\documentclass[12pt, draftclsnofoot, journal, letterpaper, onecolumn]{IEEEtran}
\IEEEoverridecommandlockouts

\usepackage{amsfonts}
\usepackage[dvips]{graphicx}
\usepackage{times}
\usepackage{cite}
\usepackage{subfigure}
\usepackage{subfig}
\usepackage{caption}
\usepackage{amsmath,amsthm}
\usepackage{array}
\usepackage{amssymb}

\newcommand{\bs}{\boldsymbol}

\usepackage{stfloats}
\usepackage{graphicx}
\usepackage{footnote}
\usepackage{booktabs}
\usepackage{array}
\usepackage{algorithmic}
\usepackage{algorithm}
\usepackage{subeqnarray}
\usepackage{cases}
\usepackage{threeparttable}
\usepackage{color}
\usepackage{url}
\usepackage{caption}

\newtheorem{theorem}{Theorem}

\newtheorem{proposition}{Proposition}
\newtheorem{lemma}{Lemma}
\newtheorem{game}{Game}
\newtheorem{definition}{Definition}

\newtheorem{question}{Question}
\newtheorem{corollary}{Corollary}

\newtheorem{example}{Example}

\usepackage{mathtools}
\DeclarePairedDelimiter\ceil{\lceil}{\rceil}
\DeclarePairedDelimiter\floor{\lfloor}{\rfloor}

\def\BibTeX{{\rm B\kern-.05em{\sc i\kern-.025em b}\kern-.08em
    T\kern-.1667em\lower.7ex\hbox{E}\kern-.125emX}}
    
\ifodd 0
\newcommand{\rev}[1]{{\color{blue}#1}}

\newcommand{\com}[1]{\textbf{\color{red} (Comment: #1) }}
\newcommand{\comg}[1]{\textbf{\color{blue} (COMMENT: #1)}}
\newcommand{\response}[1]{\textbf{\color{blue} (RESPONSE: #1)}}
\else
\newcommand{\rev}[1]{#1}

\newcommand{\com}[1]{}
\newcommand{\comg}[1]{}
\newcommand{\response}[1]{}
\fi

\ifodd 1
\newcommand{\arx}[1]{#1}
\newcommand{\arxv}[1]{}
\else
\newcommand{\arx}[1]{}
\newcommand{\arxv}[1]{#1}
\fi

\begin{document}

\title{Optimal and Quantized Mechanism Design\\for Fresh Data Acquisition\vspace{-8pt}}

\author{
 \IEEEauthorblockN{Meng Zhang, Ahmed Arafa, Ermin Wei, and Randall A. Berry}\\
     \thanks{M. Zhang, E. Wei, and Randall A. Berry are with the Department of Electrical and Computer Engineering, Northwestern University, Evanston, IL 60208 (e-mail: meng.zhang@northwestern.edu; ermin.wei@northwestern.edu; rberry@northwestern.edu). A. Arafa is with the Department of Electrical and Computer Engineering, University of North Carolina at Charlotte, Charlotte, NC 28223 (e-mail: aarafa@uncc.edu).
     }
     \vspace{-20pt}
 }   

{\maketitle}

\thispagestyle{empty}

\begin{abstract}
The proliferation of real-time applications has spurred much interest in  data freshness, captured by the {\it age-of-information} (AoI) metric. When strategic data sources have private market information, a fundamental economic challenge is how to incentivize them to acquire fresh data and optimize the age-related performance. In this work, we consider an information update system in which a destination acquires, and pays for, fresh data updates from multiple sources. The destination incurs an age-related cost, modeled as a general increasing function of the AoI. Each source is strategic and incurs a sampling cost, which is its private information and may not be truthfully reported to the destination. The destination decides on the price of updates, when to get them, and who should generate them, based on the sources' reported sampling costs. We show that a benchmark that naively trusts the sources' reports can lead to an arbitrarily bad outcome compared to the case where sources truthfully report. 
To tackle this issue, we design an  optimal (economic) mechanism for timely information acquisition following Myerson's seminal work. 
To this end, our proposed optimal mechanism minimizes the sum of the destination's age-related cost and its payment to the sources, while ensuring that the sources truthfully report their private information and will voluntarily participate in the mechanism. 
However, finding the optimal mechanisms may suffer from \textit{prohibitively expensive computational overheads} as it involves solving a nonlinear infinite-dimensional optimization problem. 
We further propose a quantized version of the optimal mechanism that achieves asymptotic optimality, maintains the other economic properties, and enables one to tradeoff between optimality and computational overheads.
 Our analytical and numerical studies show that (i) both the optimal and quantized mechanisms can lead to an unbounded benefit under some distributions of the source costs compared against a benchmark; (ii) the optimal and quantized mechanisms are most beneficial when there are few sources with heterogeneous sampling costs.
 \end{abstract}

\section{Introduction}



 The rapidly growing  number of mobile devices and the dramatic increase in real-time applications has driven interest in fresh data as measured by the \textit{age-of-information} (AoI) \cite{AoI2}. {Real-time applications in which 
fresh data is critical include real-time monitoring, data analytics, and vehicular networks. For example, real-time knowledge of traffic information and the speed of vehicles is crucial in autonomous driving and unmanned aerial vehicles.
Another example is 
 real-time mobile crowd-sensing (or mobile crowd-learning \cite{AoIEcon3}) applications, in which  a platform is fueled by mobile users' participatory contribution of real-time data.
This class of examples  includes
 real-time traffic congestion and accident information on Google Waze \cite{Waze} and real-time location information for scattered commodities and resources (e.g., GasBuddy \cite{GasBuddy}).}
 
 
 
Keeping data fresh relies on frequent data generation, processing, and sampling, which can lead to significant (sampling) costs for the data source. In practice, data sources (i.e., fresh data contributors) are \textit{self-interested} in the sense that they may have their own interests different from those of data destinations  (i.e., fresh data requestors). Consequently, the participation of sources relies on proper incentives from the destination. The resulting economic interactions between sources and destinations constitute \textit{fresh data markets}, which have been studied in \cite{AoIEcon1,AoIEcon2,AoIEcon3,AoIEcon4}.





The existing studies on fresh data markets in \cite{AoIEcon1,AoIEcon2,AoIEcon3,AoIEcon4} designed incentives assuming complete information. 
A crucial economic challenge not addressed in these works is dealing with {\it market information asymmetry}. 
Specifically, sources in practice may have private (market) information (e.g., sampling cost and data freshness) that is unknown by others.
Therefore, they may manipulate the outcome of the system (e.g., their subsidies and the scheduling policies) by
misreporting such private information to their own advantages. 
To the best of our knowledge, no existing work has addressed 
fresh data markets with such asymmetric information. 
Motivated by the above issue, this work aims to solve the following key question:
\begin{question}
	How should a destination acquire fresh data
	from self-interested sources with market information asymmetry?
\end{question}

\subsection{Challenges and Solution Approach}


Existing related studies on information asymmetry in data markets  (without considering data freshness) have identified two different levels of possible manipulation \cite{IA2,IA3,IA4,IA5,DM1,IV}, depending on whether data is \textit{verifiable}, i.e., whether the destination can verify the authenticity (or freshness) of data. These
two levels of manipulation are:
\begin{enumerate}
    \item \textit{Market information misreporting.} 
    For \textit{verifiable data}, a source  may benefit from misreporting its cost and quality information (as in, e.g., \cite{IA2,IA3,IA4,IA5,DM1}).
    \item \textit{Data fraud.}
    For \textit{unverifiable data}, a source  may even fake the data itself, e.g., by sending dummy data
    to avoid incurring corresponding costs (as in, e.g., \cite{IV}).
\end{enumerate}

As a first step towards tackling a fresh data market with asymmetric information, this work focuses on the first type of manipulation due to  misreporting private cost information and assumes \textit{verifiable fresh data.}
Even this  level of misreporting is challenging and  may lead to an arbitrarily bad loss, as we will analytically show in Section \ref{Illustrate}.

\rev{In the economics literature, a standard approach for designing markets with asymmetric information is via the
{\it optimal mechanism design} approach of Myerson \cite{Myerson}. 
Many standard  optimal mechanism design problems are linear and can be reduced to computing a ``posted price'', which is computationally efficient (e.g., \cite{Myerson}). Different from the standard setting, our fresh data market framework features a non-linear age-related cost. This nature of AoI requires a new design of optimal mechanisms and problem formulations. Once formulated, finding
the optimal mechanisms may suffer from \textit{prohibitively expensive computational overheads} as it involves solving a nonlinear infinite-dimensional optimization problem due to the age-related cost.}

To this end, we leverage the optimal mechanism design approach to optimize an AoI-related performance and address
the following question:
\begin{question}
How should a destination design a computationally efficient and optimal mechanism for acquiring fresh data?
\end{question}

We summarize our contributions as follows:
\begin{itemize}
	\item \emph{Fresh Data Market Modeling with Private Cost Information.} 
	We develop a new analytical model for a fresh data market with  private cost information and allow multiple sources to strategically misreport this information.
	To the best of our knowledge, this is the first work in the AoI literature to address market information asymmetry. 
%
%
%

\item {\emph{Optimal Mechanism Design.} 
{We first show that the optimality of a special and simplified class of mechanisms,} based on which we then transform the formulated problem into the optimal mechanism design problem.
	  The infinite-dimensional nonlinear nature of
	    the problem makes it different from the standard setting. We then solve the problem using tools from infinite dimension functional optimization and analytically derive the  optimal solution.
\item \emph{Quantized Mechanism Design.} To further reduce computational overheads, we design a
quantized mechanism while maintaining the sources' truthfulness. 
This achieves asymptotic optimality and enables one to make tradeoffs between optimality and computational overhead  by tuning the quantization step size.}


\item \emph{Performance Comparison.} 
Our analytical and numerical results show that when the sampling cost is exponentially distributed, the performance gains of our optimal mechanism can be unbounded compared against a benchmark mechanism. In addition,
the optimal mechanism is most beneficial when there are fewer sources with more heterogeneous sampling costs.

\end{itemize}

We organize the rest of this paper as follows. In Section \ref{Relate}, we discuss some related work. In Section \ref{Sysm}, we describe the system model and the mechanism design problem formulation. In Sections \ref{Single} and \ref{Multi}, we develop the optimal mechanisms for single-source systems and multi-source systems, respectively. In Section \ref{Quanti}, we develop the quantized mechanism. Section \ref{Gen} studies the optimal mechanism design under general virtual cost functions, which will be defined in Sections \ref{Single} and \ref{Multi}. 
We provide some analytical and numerical results in Section \ref{PC} to evaluate the performances of the optimal mechanism and the quantized mechanism, and we conclude the paper in Section \ref{Conclusion}. Due to the space limit, most proofs

\section{Related Work}\label{Relate}

\textbf{Age-of-Information:} {The AoI metric has been introduced and analyzed in various contexts in the recent years (e.g., \cite{AoI2,AoI4,New1,New2,AoI7,energy1,AoI12,energy2,NewAge1,NewAge2,NewAge3}). Of particular relevance to this work are those pertaining to the economics of fresh data \cite{AoIEcon1,AoIEcon2,AoIEcon3,AoIEcon4}. The most closely-related studies to ours are in \cite{AoIEcon2,AoIEcon3}, which consider systems with destinations using dynamic pricing schemes to incentivize sensors to provide fresh updates. The sources in \cite{AoIEcon2,AoIEcon3} are {\it 
price-taking}, i.e., the sources simply optimize their current payoffs given the current price and do not anticipate how this may impact future prices or decisions.  In our case, the sources are {\it strategic}, i.e., they aim to maximize their longer term payoffs. More importantly,
none of this prior work has considered the role of private market information as we do here.}

\textbf{Optimal Mechanism Design:}
There exists a rich economics literature on  optimal mechanism design (e.g., \cite{Myerson,mechanism2,mechanism3,mechanism4,mechanism5,mechanism6}). Our approach is based on Myerson's characterization of incentive compatibility and optimal mechanism design \cite{Myerson}.
{In particular, our setting is similar to a line of work on} optimal procurement mechanism (also known as reverse auction) design (e.g., \cite{mechanism2,mechanism3,mechanism4,mechanism5,mechanism6}), in which a buyer designs a mechanism for purchasing items from multiple suppliers and revealing their private quality information (as opposed to the more common case where a mechanism is used to sell items to multiple buyers).
However, existing mechanisms cannot be directly applied here due to differences in the problem setting induced by the age-related cost functions (e.g., linear programming in \cite{mechanism2,mechanism3,mechanism4,mechanism5} and combinatorial optimization in \cite{mechanism6}).

\textbf{Approximately Optimal Mechanism Design:}
Another closely related direction is approximately optimal mechanism design (e.g., \cite{mechanism1,approx1,approx2,approx3,approx4,approx5} and surveys in \cite{survey1,survey2}). Approximate mechanisms have been proposed to deal with a wide range of practical issues such as  bounded communication overheads (e.g., \cite{approx4,approx5,approx6}), bounded computational overheads (e.g., \cite{approx1,approx2,approx3}), and limited distributional knowledge (e.g., \cite{mechanism1}). In particular, \cite{approx5,approx6} designed quantized mechanisms, quantizing  the infinite-dimensional space of agents' reporting strategies
for reducing communication overheads. On the other hand, references \cite{approx1,approx2,approx3} mainly proposed approximate mechanisms to reduce the computational overheads for combinatorial problems, which is not the case here. 
Our quantized mechanism differs from these mechanisms in that it
aims at reducing computational overheads due to the underlying nonlinear infinite-dimensional optimization.

\textbf{Information Acquisition:}
There has been a recent line of work on viewing data as an economic good.
A growing amount of attention has been placed on understanding the interactions between the strategic nature of data holders and the statistical inference and learning tasks that use data collected from these holders (e.g., \cite{IA2,IA3,IA4,IA5,DM1,IV}). 
In this line of research, a data collector designs mechanisms with payments to incentivize data holders to reveal data, under private information. However, none of the studies in this line of research has  considered data freshness.

 \section{System Model and Problem Formulation}\label{Sysm}
 \subsection{System Model}

We consider an information update system in which a set $\mathcal{I}=\{1\leq i\leq I\}$ of data sources  (such as Internet-of-Things devices) generate data packets and send them to one destination.

 {The destination is interested in incentivizing the sources to generate and transmit fresh data packets by subsidizing the contributing sources for each update. The destination thus aims to trade off its age-related cost and the monetary cost of the subsidies.
		Each source $i$ incurs a sampling cost $c_i$ for each fresh data update it generates and transmits. Hence, each source $i$ aims to trade off its payment, sampling cost, and the update frequency.}

 \subsubsection{Data Updates and Scheduling} \label{DUS}
 We consider a generate-at-will model (as in, e.g., \cite{AoI7,AoI12}), in which the sources are able to generate and send a new update when requested by the destination. {We assume  instant update arrivals at the destination, with negligible transmission delay (as in, e.g., \cite{AoI12}). }

The destination's data acquisition policy consists of two decision sets, namely \textit{the update policy} $\mathcal{X}$ and the (source) \textit{scheduling policy} $\mathcal{S}$.
In particular, the 
update policy requested by the destination determines 
a sequence of times to request updates given by
$\mathcal{X}\triangleq\{x_{k}\}_{k\in\mathbb{N}}$,
where every $x_{k}\geq 0$ denotes the interarrival time between the $(k-1)$-th and $k$-th updates. The \textit{scheduling policy} $\mathcal{S}\triangleq\{s_{i,k}\}_{i\in\mathcal{I},k\in\mathbb{N}}$ is a set of binary indicators specifying which source is to be selected to generate the $k$-th update. That is, $s_{i,k}=1$ indicates that source $i$ is selected for the $k$-th update and $s_{i,k}=0$ indicates otherwise. The scheduling policy $\mathcal{S}$ should satisfy, $\forall k\in\mathbb{N}$,
\begin{align}
   { \sum_{i\in\mathcal{I}} s_{i,k}=1,~s_{i,k}\in\{0,1\},} \label{Con-1}
\end{align}
i.e., at each update, exactly one source is to be selected.

Let $y_{i,\kappa}$ denote the interarrival time between $\kappa$-th and $(\kappa-1)$-th updates generated by source $i$. Mathematically,
\vspace{-0.5cm}
\begin{align}
    y_{i,\kappa}=\sum_{k=j(i,\kappa-1)+1}^{j(i,\kappa)} x_{k}, \forall i\in\mathcal{I}, \kappa\in\mathbb{N},
\end{align}
where $j(i,\kappa)$ indicates that the $j(i,\kappa)$-th update received by the destination is the $\kappa$-th update generated by source $i$, i.e., $\sum_{k=1}^{j(i,\kappa)}s_{i,k}=\kappa$ and $\sum_{k=1}^{j(i,\kappa)-1}s_{i,k}=\kappa-1$ for all $i\in\mathcal{I}$ and $\kappa\in\mathbb{N}$.

  Each source $i$'s data updates are subject to a maximal update frequency constraint (as in \cite{AoI7}), given by
  \vspace{-0.5cm}
\begin{align}
    \limsup_{K\rightarrow \infty} \frac{\sum_{\kappa=1}^{K} y_{i,\kappa}}{K}\geq \frac{1}{f_{i,\max}},~\forall i\in\mathcal{I}, \label{Con-2}
\end{align}
where $f_{i,\max}$ is the maximal allowed average update frequency for source $i$, 
which could reflect constraints on the resources available to this source (e.g., CPU power).

 \subsubsection{Age-of-Information}  The Age-of-Information (AoI) at time $t$ is  defined as \cite{AoI2}
 \vspace{-0.1cm}
 \begin{align}
     \Delta_t(\mathcal{X})=t-U_t,
 \end{align}
  where $U_t$ is the time stamp of the most recently received update before time $t$, i.e., 
  \begin{equation}
      U_t=\max_{k\in\mathbb{N}\cup\{0\}} \sum_{j=0}^k x_{j}\quad {\rm s.t.}\quad \sum_{j=0}^k x_{j}\leq t,
  \end{equation}
and we define $x_0\triangleq 0$.


\subsubsection{Source's Sampling Cost and Private Information}
We denote the 
source $i$'s unit \textit{sampling cost} by $c_i$ for each update, which is its \textit{private information}. We consider a Bayesian setting in which each source $i$'s sampling cost is drawn from  $\mathcal{C}_i=[\underline{c}_i,\bar{c}_i]$. We define $\mathcal{C}=\prod_{i\in\mathcal{I}}\mathcal{C}_i$. Let $\Gamma_i(c_i)$ be the cumulative distribution function (CDF) and $\gamma_i(c_i)$ be the probability density function (PDF) for source $i$; we assume that only source $i$'s prior distribution is known by the destination and sources other than $i$.\footnote{In the case where such distributional knowledge is unavailable, one can further consider  prior-free approximately optimal mechanism design, as in  \cite{mechanism1}, which will be left for future work.}

 \subsubsection{Destination's AoI Cost} 
 We introduce an AoI cost function $g(\Delta_t(\mathcal{X}))$ to represent the destination's level of dissatisfaction for data staleness. We model it as a general \textit{non-negative} and \textit{increasing} function in $\Delta_t(\mathcal{X})$. 
We can specify the AoI cost function based on applications. For instance, in online learning (e.g., advertisement placement and online web ranking \cite{Online1,Online2}), 
one can 
use $g(\Delta_t(\mathcal{X}))=\Delta_t^\alpha$ with $\alpha \geq 0$.

 We further define the destination's cumulative AoI cost as
  \vspace{-0.1cm}
\begin{align}
    G(x)\triangleq\int_0^xg(\Delta_t)d \Delta_t,
\end{align}
which denotes the aggregate cost for an interarrival time $x$.
Note that $G(x)$ is convex in $x$ since $G''(x)=g'(x)>0$.

 \subsection{Mechanism Design and Reporting Game}

\begin{figure}[t]
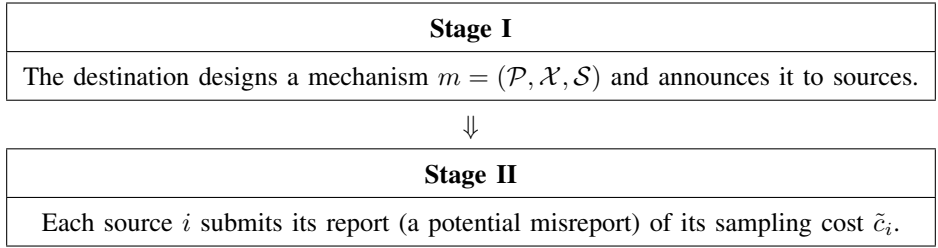

	\centering
	\small{
	\begin{tabular}{c}
		\hline
		\multicolumn{1}{|c|}{ \textbf{Stage I}}        \\ \hline
		\multicolumn{1}{|c|}{The destination designs a mechanism $m=(\mathcal{P},\mathcal{X},\mathcal{S})$ and announces it to sources.} \\ \hline
		$\Downarrow   $                                  \\ \hline
		\multicolumn{1}{|c|}{\textbf{Stage II}}        \\ \hline
		\multicolumn{1}{|c|}{Each source $i$ submits its report (a potential misreport) of its sampling cost $\tilde{c}_i$.}         \\ \hline
	\end{tabular}
	\caption{The two-stage interaction between the destination and the sources.}	\label{Stack1}}
	\vspace{-0.6cm}
\end{figure}

Fig. \ref{Stack1} depicts the interaction between the destination and the sources:
the destination in Stage I designs an (economic) mechanism for acquiring each source's report of its sampling cost
and 
data updates; the sources in Stage II report their age coefficient $\tilde{\bs{c}}\triangleq\{\tilde{c}_i\}_{i\in\mathcal{I}}$, where $\tilde{c}_i$ denotes source $i$'s report. A mechanism takes the sources' reports (potential misreports) of their sampling costs as the input of the data acquisition policy and for determining the monetary reward to each source. Mathematically,
 a general mechanism $m=(\mathcal{P},\mathcal{X},\mathcal{S})$ is a tuple of a payment rule $\mathcal{P}$, an update policy $\mathcal{X}$, and a scheduling policy $\mathcal{S}$. 
 The prices (i.e., rewards) can be different across different updates and sources. That is,  $\mathcal{P}\triangleq\{p_{i,k}\}_{i\in\mathcal{I},k\in\mathbb{N}}$, where {$p_{i,k}: \mathcal{C}\rightarrow \mathbb{R}_+$}. The sets $\mathcal{X}$ and  $\mathcal{S}$ are defined in Section \ref{DUS}.
Policies $\mathcal{P}$, $\mathcal{X}$, and $\mathcal{S}$ are functions of the sources' reported costs $\tilde{\bs{c}}\triangleq\{\tilde{c}_i\}_{i\in\mathcal{I}}$.
 
 
 

\subsubsection{Reporting Game in Multi-Source Systems} 
When there are multiple sources (i.e., $I\geq 2$), the mechanism $m$ induces a {\it reporting game} among the sources:
\begin{game}[Reporting Game]
 The reporting game $\mathcal{G}$ is a tuple given by $\mathcal{G}=(\mathcal{I}, \{\mathcal{C}_i\}_{i\in\mathcal{I}}, \{P_i\}_{i\in\mathcal{I}})$, defined as:
 \begin{itemize}
 \item Players: the set of all sources $\mathcal{I}$;
     \item Strategy space: each source $i$'s reporting strategy is $\tilde{c}_i\in\mathcal{C}_i$;
     \item Payoff: each source $i$ has a payoff function:\footnote{\rev{The infimum limit in \eqref{Pc} implies that each source is concerned about the worse-case scenario of its payoff.}}
\begin{align}
    {P_i\left(\tilde{c}_i,\tilde{\bs{c}}_{-i},m\right)=\liminf_{K\rightarrow \infty}\frac{\sum_{k=1}^K s_{i,k}(\tilde{\bs{c}})(p_{i,k}(\tilde{\bs{c}})-c_i)}{\sum_{k=1}^{K}x_{k}(\tilde{\bs{c}})},~\forall i\in\mathcal{I}.\label{Pc}}
\end{align}
 \end{itemize}
\end{game}
  The source's payoff represents its {long-run time-average profit (its received payment minus its cost) per-unit time}.
  Note that, in related studies \cite{AoIEcon2,AoIEcon3}, the considered sources are not {\textit{strategic}}. Instead, they are assumed \textit{myopic}, i.e., not maximizing their respective long-term objectives as in \eqref{Pc}.
  
 Since each source $i$ does not know the other sources' exact sampling costs $\bs{c}_{-i}$ but
 only knows the corresponding prior distributions, a Bayesian equilibrium is \rev{induced as} \cite{GameTheory}:
 \begin{definition}[Bayesian Equilibrium]
 A Bayesian equilibrium is a sources' reporting profile $\tilde{\bs{c}}^*(m)=\{\tilde{c}_i^*(m)\}_{i\in\mathcal{I}}$ such that, for all $i\in\mathcal{I}$,\footnote{{Note that $\tilde{\bs{c}}^*_{-i}(m)$ implicitly depends on $\tilde{\bs{c}}_{-i}$.}}
 \begin{align}
     \mathbb{E}_{{\bs{c}}_{-i}}[P_i\left(\tilde{c}^*_i(m),\tilde{\bs{c}}^*_{-i}(m),m\right)]\geq   \mathbb{E}_{{\bs{c}}_{-i}}[P_i\left(\tilde{c}_i,\tilde{\bs{c}}^*_{-i}(m),m\right)],~\forall \tilde{c}_i\in\mathcal{C}_i, \label{BE}
 \end{align}
 where $\tilde{\bs{c}}_{-i}^*(m)=\{\tilde{c}_j^*(m)\}_{j\neq i}$.
 \end{definition}
 In other words, a Bayesian equilibrium depicts a strategy profile where each player maximizes its expected payoff assuming the strategy of the other players is fixed.

The destination aims to design an \textit{optimal mechanism} to minimize its expected (long-term time average) overall cost, i.e., {the sum of the expected (long-term time average) AoI cost plus the expected (long-term time average) payments to the sources}, defined as
\begin{align}
    {J(m)=\mathbb{E}_{\bs{c}}\left[\limsup_{K\rightarrow \infty}\frac{\sum_{k=1}^K[G\left(x_k (\tilde{\bs{c}}^*(m))\right)+\sum_{i\in\mathcal{I}}s_{i,k}\left(\tilde{\bs{c}}^*(m)\right)p_{i,k}\left(\tilde{\bs{c}}^*(m)\right)]}{\sum_{k=1}^{K}x_k(\tilde{\bs{c}}^*(m))}\right]},\label{Overall}
\end{align}
where $\tilde{\bs{c}}^*(m)$ is the Bayesian equilibrium defined in \eqref{BE}. 
 
{We remark that an alternative approach to the long term average is to consider the discounting model, where the destination's and the sources' objectives are discounted over time, as in \cite{AoIEcon4}. We note that such a different model may lead to a similar mechanism structure as in this work. We
present detailed analysis in Appendix \ref{DM}.}

Each source $i$ may have incentive to misreport its private information $\tilde{c}_i$. However, 
according to the {\it revelation principle} \cite{Myerson}, 
for any mechanism $m$, there exists an incentive compatible (i.e. truthful) mechanism $\tilde{m}$ such that 
$J(m)=J(\tilde{m})$. This allows us to replace all
 $\tilde{\bs{c}}^*(m)$ 
in \eqref{Overall} by $\bs{c}$,
restrict our attention to incentive compatible mechanisms, and impose the following {\textit{incentive compatibility (IC)}} constraint:
\begin{align}
    {\rm  IC}:~~ c_i\in \arg\max_{\tilde{c}_i\in\mathcal{C}_i}
    ~\mathbb{E}_{\bs{c}_{-i}}[P_i\left(\tilde{c}_i,{\bs{c}}_{-i},m\right)], \forall i\in\mathcal{I}.\label{IC}
\end{align}

 
Furthermore, a mechanism should satisfy the following  \textit{(interim) individual rationality (IR)}  constraint:
\vspace{-0.5cm}
\begin{align}
  {\rm IR}: ~~\max_{\tilde{c}_i\in\mathcal{C}_i}~
    \mathbb{E}_{\bs{c}_{-i}}[P_i\left(\tilde{c}_i,\tilde{\bs{c}}_{-i}^*(m),m\right)]\geq 0,~\forall i\in\mathcal{I}.\label{IR}
\end{align}
That is, each source should not receive a negative expected payoff; otherwise, it may choose not to participate in the mechanism.

\subsubsection{Single-Source System}
We now discuss a special case where there is only one source. Hence, we can drop the index $i$ and there exists no game-theoretic interaction among sources. The incentive compatibility and the individual rationality constraints are then reduced to:
\begin{align}
    &{\rm  IC-S}:\quad c\in \arg\max_{\tilde{c}\in\mathcal{C}}
    P\left(\tilde{c},m\right),\label{IC-S}\\
  &{\rm IR-S}:\quad \max_{\tilde{c}\in\mathcal{C}}
    P\left(\tilde{c},m\right)\geq 0.\label{IR-S}
\end{align}
 




 \subsection{Problem Formulation} 
 The destination seeks to find a  mechanism $m$ to minimize its overall cost:
 \begin{subequations}\label{Problem_main}
 \begin{align}
    \min_{m}\quad &J(m)\\
    {\rm s.t.}\quad & \eqref{Con-1}, \eqref{Con-2}, {\rm IC~in}~\eqref{IC}~({\rm or}~ \eqref{IC-S})~{\rm and~IR~in}~\eqref{IR}~({\rm or}~\eqref{IR-S}).
 \end{align}
\end{subequations}
{\it This is potentially a challenging optimization problem as the space of all mechanisms is infinite dimensional and further the constraints in \eqref{IC} and \eqref{IR} are non-trivial.}

We will now show that a special, simplified, class of $m$ satisfying \eqref{IC} and \eqref{IR} is optimal.
 \begin{definition}[Equal-Spacing and Flat-Rate Mechanism]\label{Def1}
A mechanism $m=(\mathcal{P},\mathcal{X},\mathcal{S})$ is \textbf{equal-spacing} and \textbf{flat-rate} if 
\begin{align}
    p_{i,k}(\cdot)=p_i(\cdot)~{\rm and}~x_k(\cdot)=x(\cdot)~,\forall k\in\mathbb{N}, i\in\mathcal{I},\label{esfp}
\end{align}
for some functions $p_i: \mathcal{C}\rightarrow \mathbb{R}_+$ and $x: \mathcal{C}\rightarrow \mathbb{R}_+$.
\end{definition}
 \begin{definition}[(Randomized) Stationary Scheduling]\label{Def5}
The scheduling policy $\mathcal{S}$ is said to be stationary if, for all $i\in\mathcal{I}$, 
we have that given any $\bs{c}$, $s_{i,k}(\bs{c})$ is chosen randomly at each time $k$ and is \textbf{independent and identically distributed} (i.i.d) across $k$ and satisfies
\begin{align}
    {\rm Pr}( s_{i,k}(\tilde{\bs{c}})=1)=\pi_i(\tilde{\bs{c}}),~\forall \tilde{\bs{c}}\in\mathcal{C}, k\in\mathbb{N}, i\in\mathcal{I},\label{prob}
\end{align}
for some functions $\pi_i: \mathcal{C}\rightarrow [0,1]$ satisfying $\sum_{i\in\mathcal{I}}\pi_i(\bs{\tilde{c}})=1$.
 \end{definition}
 The stationary scheduling policies defined above are memoryless, in the sense that $s_{i,k}$ are independent across time. 
 We now introduce the following lemma which shows that the existence of optimal mechanisms with these properties:
\begin{lemma}\label{L1}
There exists an optimal mechanism $m^*=(\mathcal{P}^*,\mathcal{X}^*,\mathcal{S}^*)$ that is (i) equal-spacing and flat-rate, satisfying Definition \ref{Def1}, and (ii) its scheduling policy $\mathcal{S}^*$ is stationary, satisfying Definition \ref{Def5}.
\end{lemma}
\arx{We present the proof of Lemma \ref{L1} in Appendix \ref{ProofL1}.}
\arxv{\textit{Due to space limits, we present the detailed proofs of all lemmas and  theorems in our online technical report \cite{technical}.}}
The proof of Lemma \ref{L1} involves showing that, for any optimal mechanism $m^*$, we can always construct an equal-spacing and flat-rate mechanism with a stationary scheduling policy that yields at most the same objective value. 
\rev{This is mainly done by leveraging the convexity of $G(\cdot)$.
Lemma \ref{L1} allows us to restrict our attention to simple mechanisms so that we can now:
\begin{enumerate}
  \item  drop the index $k$ in $p_{i,k}$ and $x_k$;
    \item generate $s_{i,k}$ according to some i.i.d. distributions (across $k$) characterized by $\pi_i$ as in \eqref{prob}.
\end{enumerate}
}
Therefore, we use $m=(\bs{p},x,\bs{\pi})$ where $\bs{p}$ is the payment profile (i.e., $\bs{p}\triangleq\{p_i\}_{i\in\mathcal{I}}$), and $\bs{\pi}$ is the probability profile (i.e., $\bs{\pi}\triangleq\{\pi_i\}_{i\in\mathcal{I}}$).
It follows that, under an equal-spacing and flat-rate mechanism with a stationary scheduling policy, each source $i$'s payoff in \eqref{Pc} becomes:
{\begin{align}
     P_i\left(\tilde{c}_i,\tilde{\bs{c}}_{-i},m\right)=\frac{\pi_i(\tilde{\bs{c}})(p_i(\tilde{\bs{c}})-c_i)}{x(\tilde{\bs{c}})},~\forall i\in\mathcal{I}.
\end{align}}
The destination's overall cost in \eqref{Overall} becomes:
{\begin{align}
    J(m)=\mathbb{E}_{\bs{c}}\left[\frac{G\left(x (\tilde{\bs{c}}^*(m))\right)+\sum_{i\in\mathcal{I}}\pi_i\left(\tilde{\bs{c}}^*(m)\right) p_i\left(\tilde{\bs{c}}^*(m)\right)}{x(\tilde{\bs{c}}^*(m))}\right].
\end{align}}
That is, we can drop the infimum/supremum limits in \eqref{Pc} and \eqref{Overall}. 





\subsection{Naive Mechanism}\label{Illustrate}
In this subsection, we introduce a {\it naive mechanism} that satisfies Definition \ref{Def1} for single-source systems. We use this to show that such a mechanism can lead to an arbitrarily large cost for the destination when 
 $g(x)=x^\alpha$, $\alpha>0$.

\begin{example}[Naive Mechanism] Under the naive mechanism $m_N=(p_N,x_N)$,
the destination subsidizes the source's reported cost; the update policy rule $x_N(\tilde{c})$ aims at minimizing its overall cost in \eqref{Overall},
naively assuming the source's report is truthful:
\begin{align}
p_N(\tilde{c})=\tilde{c},~~{\rm and}~~x_N(\tilde{c})=\arg\min_{x\geq 0} \frac{x^{\alpha+1}/(\alpha+1)+p_N(\tilde{c})}{x}.\label{NaiveEq19}
\end{align}
 \end{example}
Solving \eqref{NaiveEq19} further gives
 \begin{align}
x_N(\tilde{c})=\left[\left(1+\frac{1}{\alpha}\right)\cdot\tilde{c}\right]^{\frac{1}{1+\alpha}}.\label{xn}
\end{align}
Given this naive mechanism, the source solves the following reporting problem:
\begin{align}
    \tilde{c}^*=\arg \max_{\tilde{c}\in\mathcal{C}}~\frac{\tilde{c}-c}{\left[\left(1+\frac{1}{\alpha}\right)\cdot\tilde{c}\right]^{1/(1+\alpha)}},
\end{align}
whose solution can be shown to be $\tilde{c}^*=\bar{c}$, i.e., the optimal reporting strategy is to report the maximal possible value.\footnote{Note that it is not immediate that a source would always report its maximum value under such a mechanism. Though a larger reported value leads to a larger payment per update, it also leads to a larger inter-arrival time between updates.} This makes the destination's overall cost to be given by
\begin{align}
    J(m_N)=\left[\bar{c}\left(1+\frac{1}{\alpha}\right)\right]^{\frac{\alpha}{1+\alpha}}. \label{Naive3}
\end{align}
Note that the ratio of the destination's objectives in \eqref{Naive3} under the source's optimal report and the true cost is  $\left(\frac{\bar{c}}{c}\right)^{\frac{\alpha}{1+\alpha}}$,
which can be arbitrarily large as $\bar{c}$ approaches infinity. Misreports leading to an arbitrarily large cost to the destination motivates the optimal mechanism design next.

 \section{Single-Source Optimal Mechanism Design}\label{Single}
 
 In this section, we start with a system with only one source. Therefore, we can drop the index $i$ in our notations.
 We
 use the results of Lemma \ref{L1} to reformulate \eqref{Problem_main} and characterize the IC and the IR constraints in \eqref{IC-S} and \eqref{IR-S}.
 The optimal mechanism design problem is then reduced to an infinite-dimensional optimization  problem, which we analytically solve and use to derive useful insights.
 
 \subsection{Problem Reformulation} 

Lemma \ref{L1} allows us to focus on equal-spacing and flat-rate mechanisms. 
The scheduling indicators satisfy $\pi(\tilde{c})=1$, since only one source is present. The equal-spacing and flat-rate mechanism is then reduced to $m=(p,x)$.

To further facilitate our analysis, we use 
$f(\tilde{c})$  to denote the \textit{update rate rule}  and $h(\tilde{c})$ to denote
the \textit{payment rate rule} such that
 \begin{align}
     h(\tilde{c})\triangleq\frac{p(\tilde{c})}{x(\tilde{c})}~{\rm and}~ f(\tilde{c})\triangleq\frac{1}{x(\tilde{c})},~\forall \tilde{c}\in\mathcal{C}. \label{f}
 \end{align}
Since \eqref{f} defines a one-to-one mapping between $(p,x)$ and $(f,h)$, we can focus on $m=(f,h)$ in the following and then derive the optimal $(p^*,x^*)$ based on the optimal $(f^*,h^*)$. 
  
    	\begin{figure}[t]
  	\centering
		\subfigure[]{\includegraphics[scale=.26]{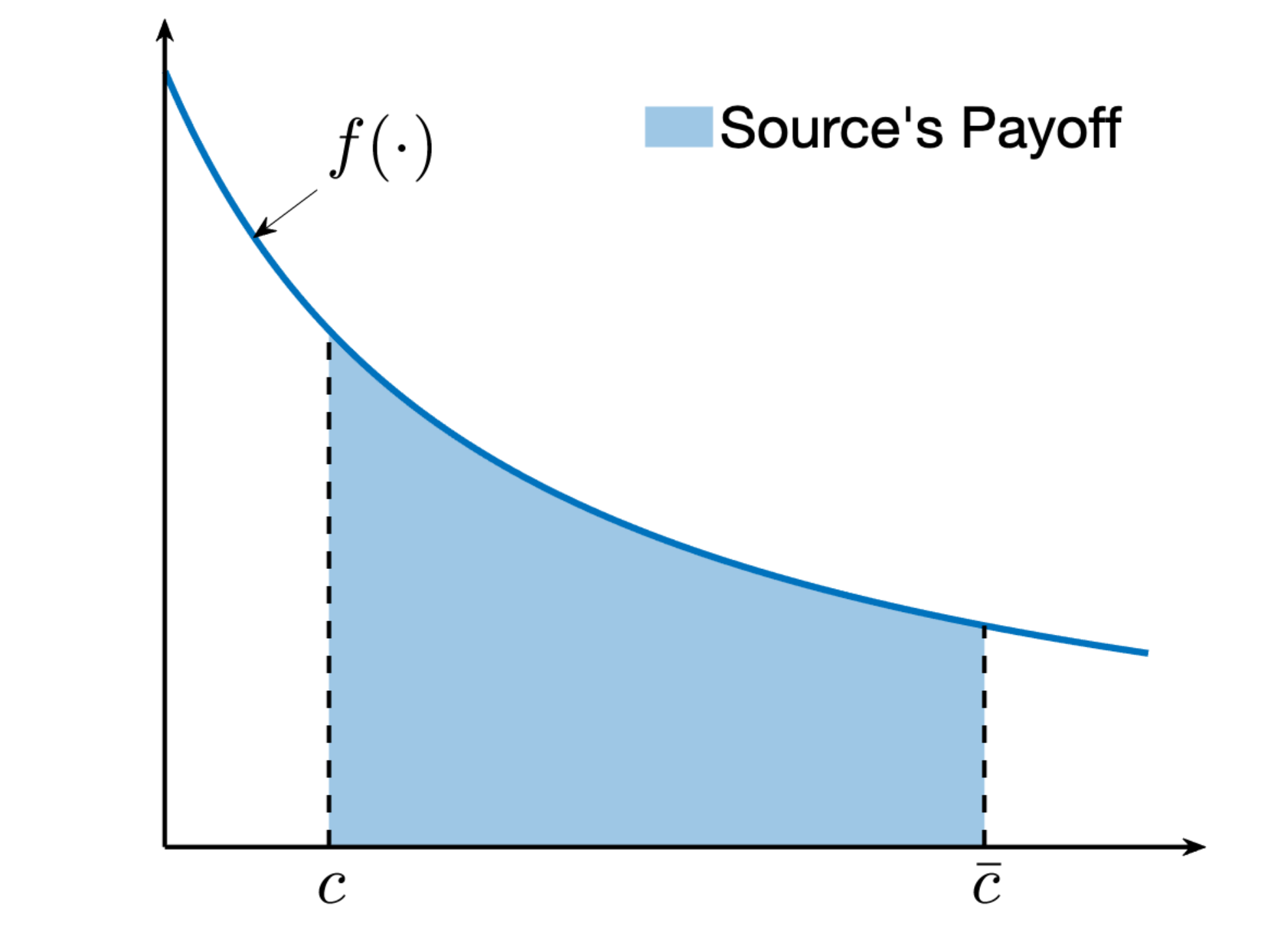}}
		\subfigure[]{\includegraphics[scale=.26]{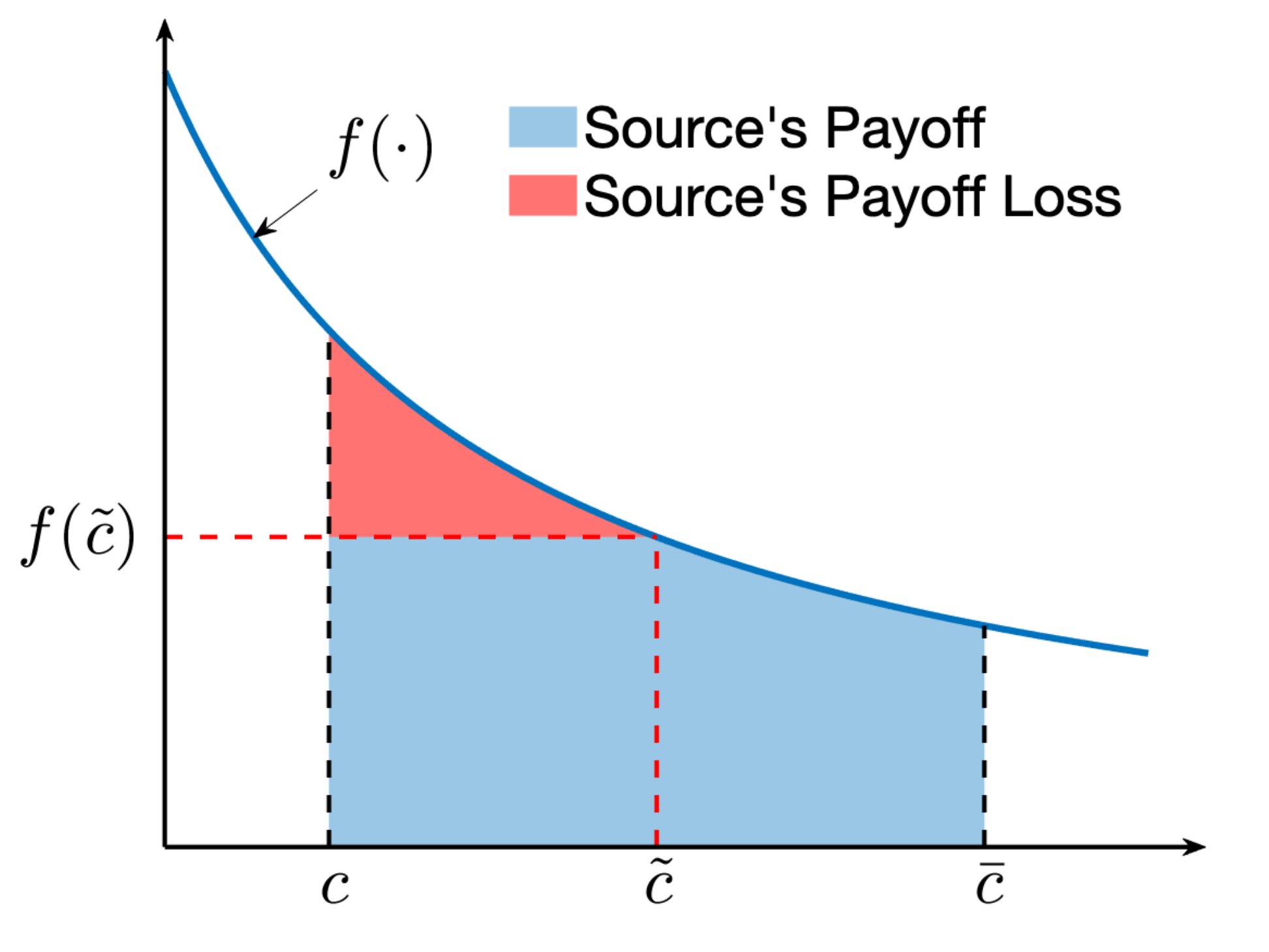}}
		\vspace{-0.3cm}
		\caption{Illustration of IC and IR under a mechanism satisfying \eqref{ICC} and \eqref{IRC}: The source's payoff comparison between (a) a truthful report ($\tilde{c}=c$) and (b) an over-report ($\tilde{c}>c$). }\label{ill}
		\vspace{-0.6cm}
	\end{figure}

  \subsection{Characterization of IC and IR}
   \subsubsection{Incentive Compatibility} We can characterize the IC constraint in  \eqref{IC} based on Myerson's work \cite{Myerson}.
  \begin{theorem}\label{T1}
 A mechanism $m=(f,h)$ is incentive compatible \textit{if and only if} the following two conditions are satisfied:
\begin{enumerate}
  \item $f(c)$ is non-increasing in $c\in\mathcal{C}$;
    \item $h(c)$ has the following form:
    \begin{align}
        h(c)=c\cdot f(c)-\int_{\underline{c}}^cf(z)dz  +A, \label{ICC}
    \end{align}
    for some constant $A\in\mathbb{R}$ (here, $A$ does not depend on $c$ but may depend on $f(\cdot)$.)
\end{enumerate}
 \end{theorem}
\arx{We present the proof of Theorem \ref{T1} in Appendix \ref{ProofT1}.}

\subsubsection{Individual Rationality}
Given an arbitrary incentive compatible mechanism satisfying \eqref{ICC}, to further satisfy the IR constraint in \eqref{IR}, we have that the minimal $A$ in \eqref{ICC} for the incentive compatible mechanism in Theorem in \ref{T1} is 
 \begin{align}
 A=\int_{\underline{c}}^{\bar{c}}f(z)dz. \label{IRC}
 \end{align}
 We will assume that this choice of $A$ is used in the following.

	
 We present an example in Fig. \ref{ill} to illustrate \eqref{ICC} and \eqref{IRC}. Under a non-increasing $f(\cdot)$ and $h(\cdot)$ satisfying \eqref{ICC} and \eqref{IRC},
 a truthfully reporting source receives a payoff of $\int_{c}^{\bar{c}}f(t)dt$, as shown in Fig. \ref{ill} (a); when the source reports $\tilde{c}$, its payoff is $(\tilde{c}-c)f(\tilde{c})+\int_{\tilde{c}}^{\bar{c}}f(t)dt$. As shown in Fig. \ref{ill} (b), such an over-report incurs a payoff loss. Similarly, an under-report would also incur a payoff loss.
These demonstrate incentive compatibility. In addition, a truthfully reporting source's payoff is always non-negative for any $c$ and approaches $0$ when $c$ approaches $\bar{c}$, as $\lim_{c\rightarrow \bar{c}}\int_{c}^{\bar{c}}f(z)dz=0$. This demonstrates individual rationality.

\subsection{Mechanism Optimization Problem}
Based on \eqref{ICC} and \eqref{IRC},
we can focus on optimizing the update rate function $f(c)$ only in the following.
By the constraint $f(c)\in[0,f_{\max}]$, it follows that
$\int_{\underline{c}}^{\bar{c}} |f(c)|^2d\Gamma(c)<+\infty$. Therefore, the update rate function $f(\cdot)$ lives in the Hilbert space $L^2(\Gamma)$ associated to  the measure of $c$, i.e. the CDF $\Gamma(c)$.

By Theorem \ref{T1} and \eqref{IRC}, we transform the destination's problem into
\begin{subequations}\label{Problem}
\begin{align}
    &\min_{f(\cdot)}J(f)\triangleq\mathbb{E}_c\left[G\left(\frac{1}{f(c)}\right)f(c)+c\cdot f(c)+\int_{c}^{\bar{c}}f(z)dz\right]\label{Problema}\\
    &~~{\rm s.t.}~f(\cdot)\in\mathcal{F}\triangleq\left\{f(\cdot): f(c) \in [0,f_{\max}], f'(c)\leq 0,~\forall c\in\mathcal{C}\right\}. \label{Problemb}
\end{align}
\end{subequations}
In particular, the objective \eqref{Problema} comes from \eqref{IRC} and \eqref{ICC}, the constraint $f'(c)\leq 0$ comes from Theorem \ref{T1}, and $f(c) \in [0,f_{\max}]$ comes from \eqref{Con-2}.

This is a functional optimization problem. 
To derive insightful results, we first relax the constraint in \eqref{Problemb} and then show when such a relaxation in fact leads to a feasible solution $f^*(\cdot)$ (i.e., when it automatically satisfies \eqref{Problemb}).

We introduce the definition of the source's \textit{virtual cost} which is analogous to the standard definition of virtual value in \cite{Myerson}:
\begin{definition}[Virtual Cost]
The source's virtual cost is
\begin{align}\phi(c)\triangleq c+\frac{\Gamma(c)}{\gamma(c)}.
\end{align} 
\end{definition}

The virtual cost allows us to transform the destination's problem as
in the following lemma:
\begin{lemma}\label{L2-1}
The objective in \eqref{Problema} can be rewritten as 
\begin{align}
  J(f)=\mathbb{E}_c\left[G\left(\frac{1}{f(c)}\right)f(c)+f(c)\phi(c)\right].\label{virtualcc}
\end{align}
\end{lemma}
\arx{We prove Lemma \ref{L2-1} in Appendix \ref{ProofL2}, which involves changing the order of integration.}
\arxv{The proof of Lemma \ref{L2-1} involves changing the order of integration.}
If we relax the constraint $f'(c)\leq 0$, Lemma \ref{L2} makes the problem in \eqref{Problem} decomposable across every $c\in\mathcal{C}$. Each subproblem is given by
\begin{align}
    \min_{f(c)\in[0,f_{\rm max}]}G\left(\frac{1}{f(c)}\right)f(c)+f(c)\phi(c),
\end{align}
which can be solved separably. We are now ready to introduce the solution to problem \eqref{Problem}:
\begin{theorem}\label{T2}
If $\phi(c)$ is non-decreasing, the optimal mechanism $m^*=(f^*,h^*)$ satisfies \eqref{ICC}, \eqref{IRC}, and 
\vspace{-0.5cm}
\begin{align}
f^*(c)=\min\left\{f_{\max}, \hat{f}(c)\right\},~\forall c\in\mathcal{C}, \label{z3}
\end{align}
where $\hat{f}(\cdot)$
satisfies
\begin{align}
\underbrace{g\left(\frac{1}{\hat{f}(c)}\right)\frac{1}{\hat{f}(c)}-G\left(\frac{1}{\hat{f}(c)}\right)}_{\rm Marginal~AoI~Cost~Reduction} = \underbrace{\phi(c)}_{\rm Virtual~Cost},~\forall c\in\mathcal{C}. \label{z3-b}
\end{align}
\end{theorem}
\arx{We present the proof of Theorem \ref{T2} in Appendix \ref{ProofT2}.} To comprehend the above results, 
the optimal solution $f^*(c)$ solves each subproblem by the following two steps: (i) search for a $\hat{f}(c)$ that equalizes the marginal AoI cost reduction and the virtual cost $\phi(c)$ for every $c\in\mathcal{C}$; (ii) project every $\hat{f}(c)$ onto the feasible set $[0,f_{\max}]$.

To see when \eqref{z3} yields a feasible solution satisfying \eqref{Problemb},
note that there always exists a unique positive value of $\hat{f}(c)$ in \eqref{z3-b}, and so
the optimal $f^*(c)$ for each $c$ in \eqref{z3} is well defined.
In addition, if $\phi(c)$ is non-decreasing in $c$, $f^*(c)$ is non-increasing in $c$.\footnote{The condition of the virtual cost $\phi(c)$ being non-decreasing is known as the regularity condition in \cite{Myerson}.}

\begin{figure}[t]
  	\centering
	\includegraphics[scale=.3]{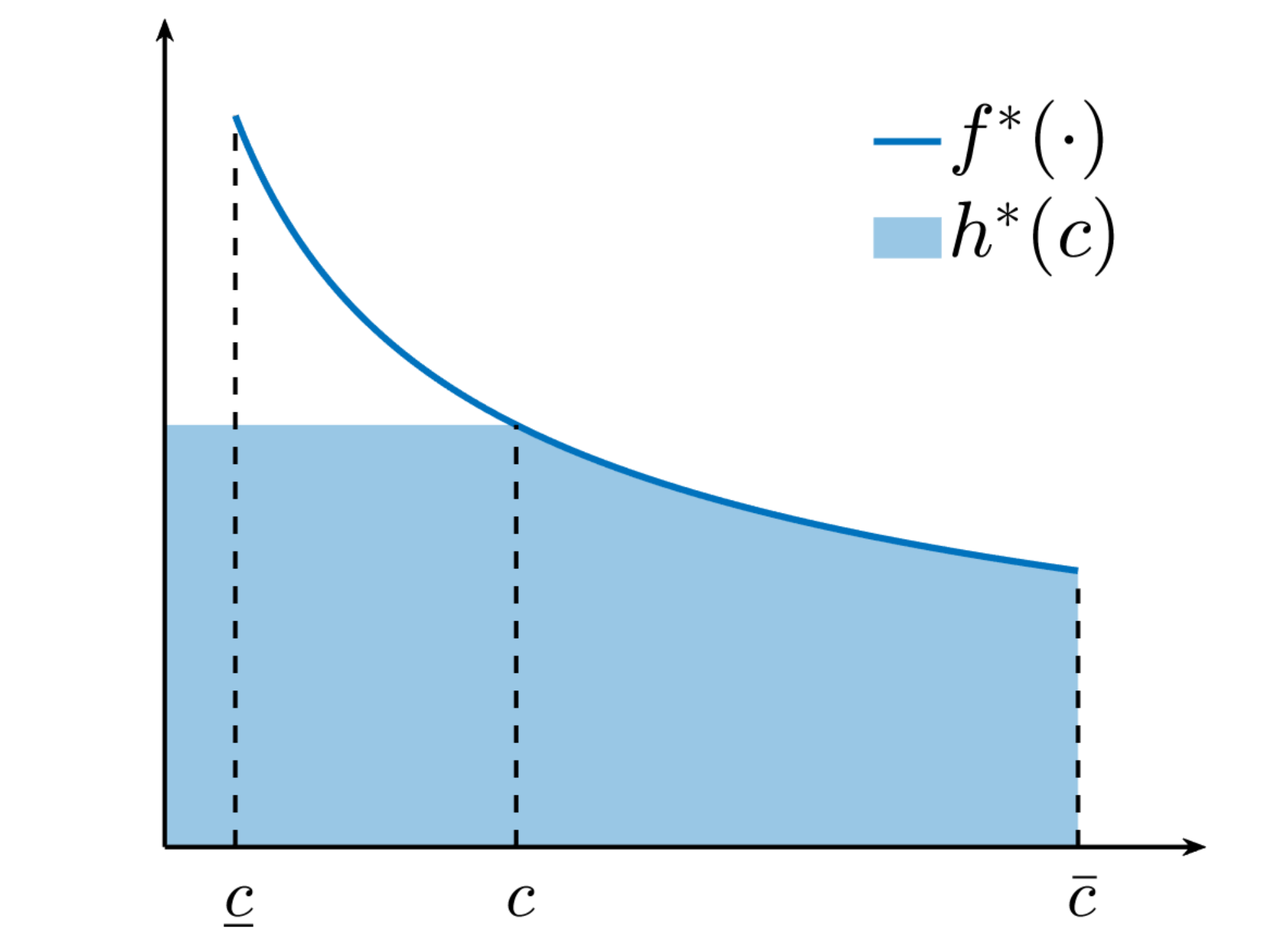}
		\vspace{-0.15cm}
		\caption{Illustration of the optimal mechanism $(f^*,h^*)$ for a single-source system. The AoI cost is $g(\Delta_t)=\Delta_t$ and the source's sampling cost follows a uniform distribution. }\label{illl-optimal}
		\vspace{-0.6cm}
	\end{figure}

A non-decreasing virtual cost $\phi(c)$ is in fact satisfied for a wide range of distributions of the source's sampling cost. Fig. \ref{illl-optimal} illustrates an example of the optimal mechanism $m^*=(f^*,h^*)$ when the source's sampling cost $c$ follows a uniform distribution. We will focus on specific distributions in Section \ref{PC} and generalize Theorem \ref{T2} to the more general (potentially not monotonic) virtual cost case in Section \ref{Gen}.

 \subsection{Differences from Classical Settings and Computational Complexity} 
 We highlight a key difference of the optimal mechanism $m^*$ satisfying Theorem \ref{T2} from some existing optimal mechanisms (e.g., in \cite{Myerson}) in classical economic settings, in which the sellers' problems can be formulated into infinite-dimensional linear programs.  Reference \cite{Myerson} showed that the optimal mechanism is a posted price mechanism in such classical settings, i.e., the optimal mechanism determines
 a posted price (equal to the virtual cost $\phi(c)$ in our case). If the source's cost is less than the posted price, then it is assigned the maximum update rate with its payment equal to the price; otherwise the source is assigned no update. 
 
 Our problem in \eqref{Problem} differs from the classical settings in the nonlinearity introduced by $G(\cdot)$,
 which brings the issue of computational complexity.
 As shown in Fig. \ref{illl-optimal}, the computation of the optimal payment rate $h^*(c)$ requires solving for $f^*(c)$ in \eqref{z3} over the entire interval $[c,\bar{c}]$, which may be computationally impractical. {We note that the computational  and the economic challenges are coupled as it requires
 the joint design with efficient computation and the satisfaction of IR and IC. In particular, standard numerical approaches\footnote{
 One example of such numerical approaches is to compute $f^*(c)$ for a discrete set of $c$ and then use the interpolation to approximate $f^*(\cdot)$.} may lead to $f$ that is not non-increasing, which violates the sufficient and necessary condition in Theorem \ref{T1}.}\footnote{This is not the case in the aforementioned existing optimal mechanism  for classical settings (e.g. in \cite{Myerson}), since the integral is reduced to the virtual cost, which is computationally efficient.}
 This motivates us to consider a computationally efficient approximation of the optimal mechanism $m^*=(f^*,h^*)$ {without affecting the satisfaction of IR and IC} in Section \ref{Quanti}.

\section{Multi-Source Optimal Mechanism}\label{Multi}

In this section, we extend our results in Section \ref{Single} to multi-source systems. The additional challenge here is that the optimal mechanism needs to take the sources' interactions with each other into account.
Similar to the single-source case, we first characterize the IC and the IR constraints, and then solve the infinite-dimensional optimization problem.

\subsection{Problem Reformulation}

 Lemma \ref{L1} allows us to focus on equal-spacing and flat-rate mechanisms with stationary scheduling policies (i.e., $m=(\bs{p},x,\bs{\pi})$). 
To further facilitate our analysis, we use 
$\bs{f}(\tilde{\bs{c}})=\{f_i(\tilde{\bs{c}})\}_{i\in\mathcal{I}}$  to denote the \textit{update rate rule}  and $\bs{h}(\tilde{\bs{c}})=\{h_i(\tilde{\bs{c}})\}_{i\in\mathcal{I}}$ to denote the \textit{payment rate rule} such that, for all $i\in\mathcal{I}$ and all $\tilde{\bs{c}}\in\mathcal{C}$:
\begin{subequations} \label{f2}
 \begin{align}
     &h_i(\tilde{\bs{c}})\triangleq\frac{\pi_i(\tilde{\bs{c}})\cdot p_i(\tilde{\bs{c}})}{x(\tilde{\bs{c}})},~f_i(\tilde{\bs{c}})\triangleq\frac{\pi_i(\tilde{\bs{c}})}{x(\tilde{\bs{c}})}, \label{f2-1}\\
    &\pi_{i}(\tilde{\bs{c}})=\frac{f_i(\tilde{\bs{c}})}{\sum_{j\in\mathcal{I}}f_j(\tilde{\bs{c}})}. \label{f2-2}
 \end{align}
 \end{subequations}
The above equations \eqref{f2} define a one-to-one mapping between $(\bs{p},x,\bs{\pi})$ and $(\bs{f},\bs{h})$.
Hence,
we can restrict our attention to $m=(\bs{f},\bs{h})$ in the following and then derive the optimal $(\bs{p}^*,x^*,\bs{\pi}^*)$. We can then generate the corresponding stationary scheduling policy $\mathcal{S}^*$ satisfying \eqref{f2-2} based on the optimal $(\bs{f}^*,\bs{h}^*)$.

 \subsection{Characterization of Incentive Compatibility and Individual Rationality}
 \subsubsection{Incentive Compatibility} 
 
 Theorem \ref{T1} can be generalized as follows to the multi-source setting to characterize the IC constraint in \eqref{IC}:

   \begin{theorem}\label{T3}
 A mechanism $m=(\bs{f},\bs{h})$ is incentive compatible \textit{if and only if} the following two conditions are satisfied:
\begin{enumerate}
  \item $f_i({c}_i,\bs{c}_{-i})$ is non-increasing in $c_i\in\mathcal{C}_i$;
    \item $h_i({c}_i,\bs{c}_{-i})$ has the following form:
    \begin{align}
        h_i(\bs{c})=c_i\cdot f_i(\bs{c})-\int_{\underline{c}_i}^{c_i}f_i(z,\bs{c}_{-i})dz  +A_i, ~\forall i\in\mathcal{I},\label{ICC-M}
    \end{align}
    for some constant $A_i\in\mathbb{R}$ for all $i\in\mathcal{I}$.
\end{enumerate}
 \end{theorem}
 \arx{We present the proof of Theorem \ref{T3} in Appendix \ref{ProofT1}.}
 \subsubsection{Individual Rationality}
Given an arbitrary incentive compatible mechanism satisfying \eqref{ICC-M}, to further satisfy the IR constraint in \eqref{IR}, we have that the minimal $A_i$ in \eqref{ICC-M} for the incentive compatible mechanism in Theorem \ref{T3} is 
 \begin{align}
 A_i=\int_{\underline{c}_i}^{\bar{c}_i}f_i(z,\bs{c}_{-i})dz,  ~\forall i\in\mathcal{I}.\label{IRC-M}
 \end{align}
 We again assume that this minimal value of $A_i$ is used in the following.
 \subsection{Mechanism Optimization Problem}

Based on  \eqref{ICC-M} and \eqref{IRC-M},
we can focus on optimizing the update rate function $\bs{f}(\bs{c})$ only in what follows, which live in the Hilbert space $L^2(\Gamma)$ associated to  the measure of $\bs{c}$.
We introduce the definition of the source $i$'s \textit{virtual cost}.
\begin{definition}[Virtual Cost]
The source $i$'s virtual cost is
\begin{align}\phi_i(c_i)\triangleq c_i+\frac{\Gamma_i(c_i)}{\gamma_i(c_i)}, ~\forall i\in\mathcal{I}.
\end{align} 
\end{definition}
The sources' virtual costs enable the problem to be transformed as in the following lemma:
\begin{lemma}\label{L2}

The destination's problem in \eqref{Problem_main} is equivalent to
\begin{subequations}\label{Problem-2b}
\begin{align}
    &\min_{\bs{f}(\cdot)}\quad J(\bs{f})\triangleq\mathbb{E}_{\bs{c}}\left[G\left(\frac{1}{\sum_{i\in\mathcal{I}}f_i(\bs{c})}\right)\sum_{i\in\mathcal{I}}f_i(\bs{c})+\sum_{i\in\mathcal{I}}\phi_i(c_i) f_i(\bs{c})\right]\label{Problema-2}\\
    &~~{\rm s.t.}\quad \bs{f}(\cdot)\in\mathcal{F}\triangleq\left\{\bs{f}(\cdot): f_i(\bs{c})\in [0,f_{i,\max}], f_i'(\bs{c})\leq 0,~\forall \bs{c}\in\mathcal{C},~i\in\mathcal{I}\right\}, \label{Problemb-2}
\end{align}
\end{subequations}
where $f_i(c_i) \in [0,f_{i,\max}]$ comes from \eqref{Con-2} and \eqref{f2}.
\end{lemma}
\arx{We prove Lemma \ref{L2} in Appendix \ref{ProofL2}, which  involves changing the order of integration.}
Different from \eqref{Problem},  the functional optimization problem in \eqref{Problem-2b} has a vector-valued function as its optimization decision.
To derive insightful results, we first omit the $f_i'(\bs{c})\leq 0$ constraints for all $i\in\mathcal{I}$, similar to our approach in Section \ref{Single}. Such constraints are automatically satisfied assuming the virtual costs are non-decreasing $\phi_i(c_i)$ as we will show later. We will extend our results to the case of general virtual costs in Section \ref{Gen}. 

To solve the problem in \eqref{Problem-2b}, we next introduce the aggregate update rate satisfying
\begin{align}
    f_{\rm agg}(\bs{c})=\sum_{i\in\mathcal{I}} f_i(\bs{c}),~\forall \bs{c}\in\mathcal{C},\label{aggregate}
\end{align}
and the following definition:
\begin{definition}[Aggregate Virtual Cost]\label{D7}
Let $\Psi(\bs{c},f_{\rm agg}(\bs{c}))$ be the aggregate virtual cost function, defined as
\vspace{-0.5cm}
\begin{subequations}\label{AVC}
\begin{align}
     \Psi(\bs{c},f_{\rm agg}(\bs{c}))=\min_{\bs{f}}\quad~& \sum_{i\in\mathcal{I}}\phi_i(c_i) {f}_i(\bs{c})\\
    ~{\rm s.t.} \quad~ &\eqref{aggregate},~f_i(\bs{c})\in [0,f_{i,\max}] ,\forall i\in\mathcal{I}. \label{AVC-c}
\end{align}
\end{subequations}
\end{definition}
The definition of the aggregate virtual cost in Definition \ref{D7} involves solving a linear programming problem parameterized by $\bs{c}$.
The intuition of solving \eqref{AVC} is as follows. Given each $\bs{c}$ and  $f_{\rm agg}(\bs{c})$, we assign the sources with higher virtual costs 
only after the sources with lower virtual costs $\phi_i(c_i)$ are fully utilized (i.e., the constraints in \eqref{AVC-c} are binding).
It is readily verified that, given $\bs{c}$, $\Psi(\bs{c},f_{\rm agg}(\bs{c}))$ is a piece-wise linear function in $f_{\rm agg}(\bs{c})$, and its differential $\partial_{f_{\rm agg}(\bs{c})}\Psi(\bs{c},f_{\rm agg}(\bs{c}))$ is a step function in $f_{\rm agg}(\bs{c})$. We now introduce the following result to further transform the destination's problem:



\begin{lemma}\label{L4}
If $\phi_i(c_i)$ is non-decreasing for all $i\in\mathcal{I}$, the destination's problem in \eqref{Problem-2b} leads to the same minimal objective value as the following problem:
\begin{subequations}\label{Problem-3}
\begin{align}
    \min_{f_{\rm agg}(\cdot)}\quad &J(f_{\rm agg})\triangleq\mathbb{E}_{\bs{c}}\left[G\left(\frac{1}{f_{\rm agg}(\bs{c})}\right)f_{\rm agg}(\bs{c})+\Psi(\bs{c},f_{\rm agg}(\bs{c}))\right]\\
    ~~{\rm s.t.} \quad &f_{\rm agg}(\bs{c})\in\left[0,\sum_{i\in\mathcal{I}} f_{i,\max}\right], ~\forall \bs{c}\in\mathcal{C}.
\end{align}
\end{subequations}
\end{lemma}
\arx{We present the proof of Lemma \ref{L4} in Appendix \ref{ProofL4}.}
 Lemma \ref{L4} transforms the vector functional optimization problem in \eqref{Problem-2b} into a scalar functional optimization problem in \eqref{Problem-3}. Therefore, after obtaining the optimal solution $f^*_{\rm agg}(\cdot)$, we can then solve the problem in \eqref{AVC} to obtain the original solution to the problem in \eqref{Problem-2b}.


	

\begin{figure}[t]
\centering
		\includegraphics[scale=.35]{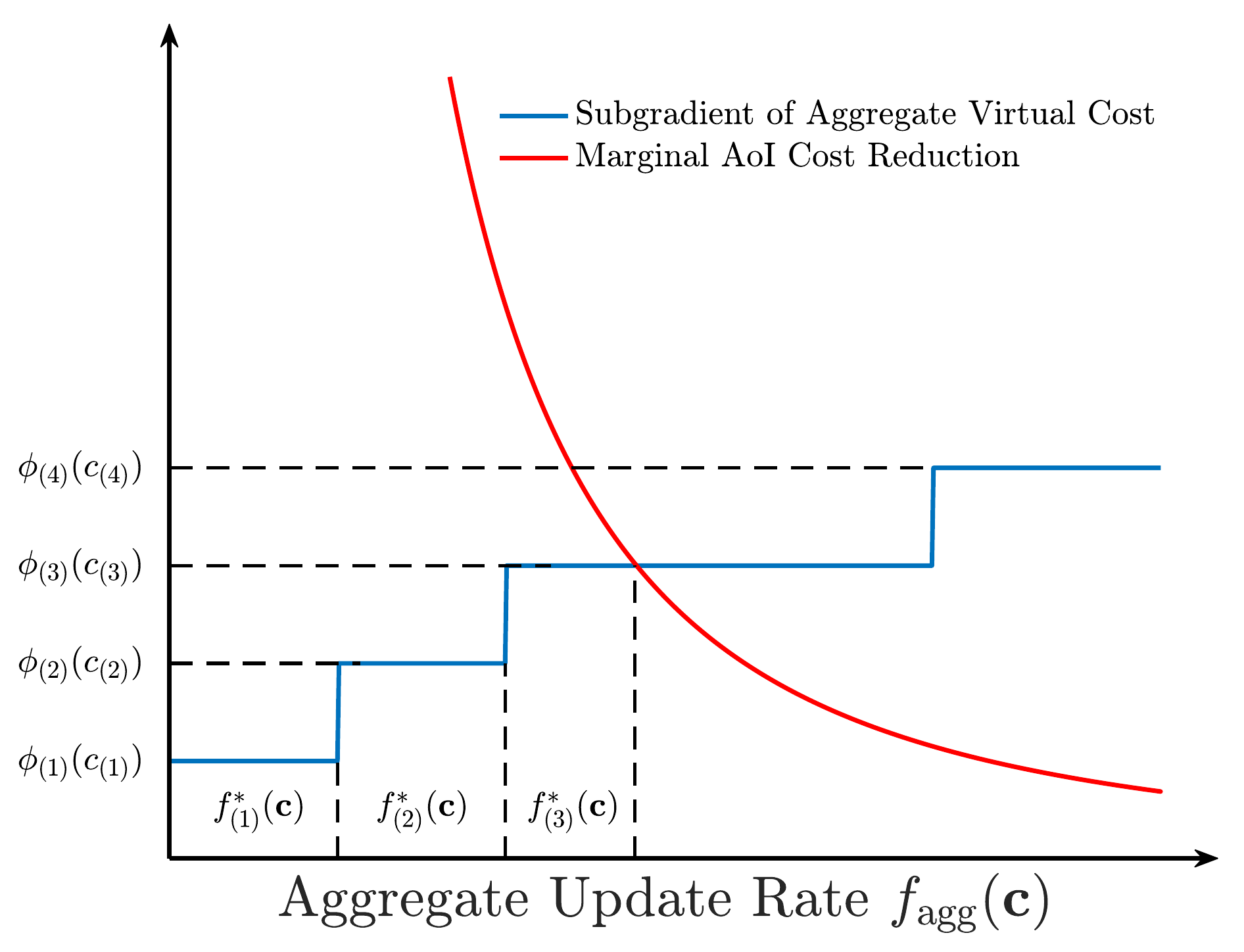}
		\vspace{-0.25cm}
		\caption{Illustration of the optimal solutions to the problems in \eqref{Problem-3} and \eqref{Problem-2b}. }\label{illlll}
		\vspace{-0.65cm}
\end{figure} 
	
We observe that the problem in \eqref{Problem-3} now becomes similar to the problem in Lemma \ref{L2} in the single-source case, with the following difference: $\Phi(\bs{c},f_{\rm agg}(\bs{c}))$ is not differentiable in $f_{\rm agg}(\bs{c}))$. Hence, 
it follows that the optimality condition of \eqref{Problem-3} can be rewritten as
\begin{align}
g\left(\frac{1}{f_{\rm agg}(\bs{c})}\right)\frac{1}{f_{\rm agg}(\bs{c})}-G\left(\frac{1}{f_{\rm agg}(\bs{c})}\right) \in \partial \Psi(\bs{c}, f_{\rm agg}(\bs{c})),~\forall c\in\mathcal{C}, \label{z4}
\end{align}
i.e., the marginal AoI cost reduction is equal to a subgradient of the aggregate virtual cost.

To understand \eqref{z4}, we first introduce the order indexing $(i)$ such that
\begin{align}
    \phi_{(1)}(c_{(1)})\leq \phi_{(2)}(c_{(2)})\leq ...\leq \phi_{(i)}(c_{(i)})\leq ...\leq \phi_{(I)}(c_{(I)}), \label{oi}
\end{align}
i.e., source $(i)$ has the $i$-th smallest virtual cost $\phi_{(i)}(c_{(i)})$.
We present an illustrative example of \eqref{z4} in Fig. \ref{illlll} for a given $\bs{c}$. As mentioned, the differential of the aggregate virtual cost in $f_{\rm agg}(\bs{c})$ corresponds to a step function, as shown in Fig. \ref{illlll}. The intersection point between the subgradient and the curve of the marginal AoI cost reduction corresponds to the solution to \eqref{z4}.



 Based on \eqref{oi} and  \eqref{z4}, we are ready to present the solution to \eqref{Problem-2b}:
\begin{theorem}\label{T4}
If the sources' virtual costs $\phi_i(c_i)$ are non-decreasing, the optimal mechanism $m^*=(\bs{f}^*,\bs{h}^*)$ satisfies \eqref{ICC-M}, \eqref{IRC-M}, and 
\begin{align}
    f_{(i)}^*(\bs{c})=\left[f_{\rm agg}^*(\bs{c})-\sum_{j=1}^{i-1}f^*_{(j)}(\bs{c})\right]_0^{f_{(i),\max}}, \forall \bs{c}\in\mathcal{C}, \forall i\in\mathcal{I}. \label{T4-eq}
\end{align}

\end{theorem}
\arx{We present the proof of Theorem \ref{T4} in Appendix \ref{ProofT4}.}
Intuitively, after obtaining the optimal aggregate update rate $f^*_{\rm agg}(\cdot)$, the problem is reduced to solving \eqref{Problem-3}. 
That is, we utilize the least expensive (in terms of virtual cost) sources first and the sources with high virtual costs are assign update rates of $0$.
Each source $(i)$'s allocated update rate $f_{(i)}^*(\cdot)$ is then the residual aggregate update rate (the aggregate update rate subtracted from assigned update rates to the first $i-1$ least expensive sources) projected onto its feasible set $[0,f_{(i),\max}]$. 

We will generalize Theorem \ref{T4} to the more general (potentially not monotonic) virtual cost case in Section \ref{Gen}, and design a computationally efficient approximation of the optimal mechanism $m^*=(\bs{f}^*,\bs{h}^*)$ in Section \ref{Quanti}.

\section{Quantized Optimal Mechanism}\label{Quanti}

We note that the optimal mechanisms (for both single-source systems and multi-source systems) may be computationally impractical, since the optimal payment rate $\bs{h}^*(\bs{c})$ for the optimal mechanisms in \eqref{ICC} and \eqref{ICC-M} require explicitly solving $\bs{f}^*(\bs{c})$  in \eqref{z3} and \eqref{T4-eq} for all $\bs{c}$.

    	\begin{figure}[t]
  	\centering
		\includegraphics[scale=.3]{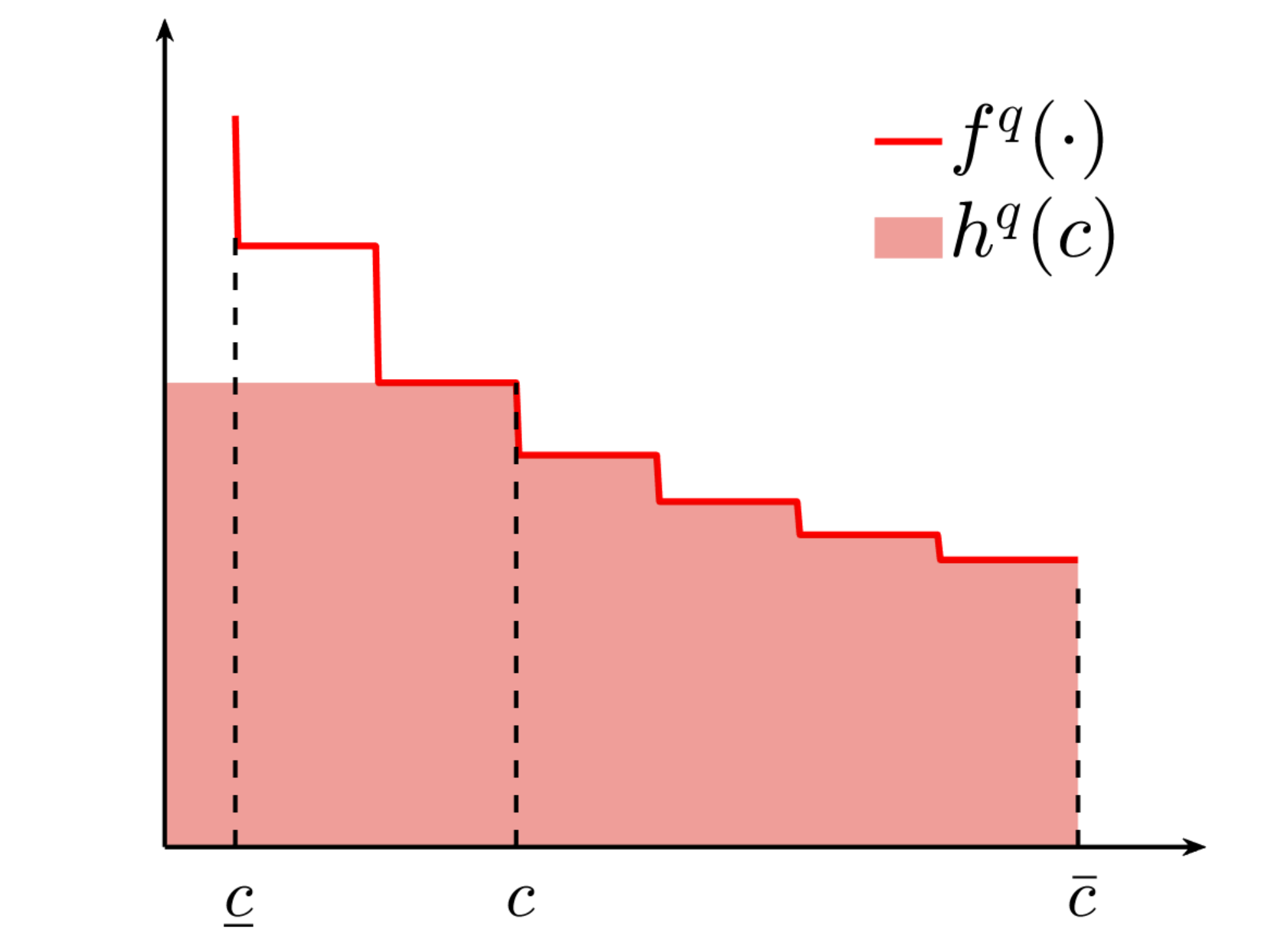}
		\vspace{-0.25cm}
		\caption{Illustration of a quantized mechanism $(f^q,h^q)$  for a single-source system. }\label{illl}
		\vspace{-0.65cm}
	\end{figure} 

Therefore, we are motivated to design a computationally efficient \textit{quantized  mechanism} that is approximately optimal while maintaining the optimal mechanism's economic properties.

\subsection{Quantized Mechanism}
\subsubsection{Quantized mechanism description}  Let $\bs{Q}(\bs{c})\triangleq\{Q_i(c_i)\}_{i\in\mathcal{I}}$ be the sources' \textit{quantized reporting profile} such that
\vspace{-0.2cm}
\begin{align}
    Q_i({c}_i)\triangleq\Delta_Q\cdot\left( \floor{\frac{c_i}{\Delta_Q}}+\frac{1}{2}\right), ~\forall i\in\mathcal{I},\label{Quantization}
\end{align}
where 
$\floor{\cdot}$ 
depicts the floor operator and $\Delta_Q$ is the quantization step size.
Based on \eqref{Quantization}, we introduce the \textit{quantized mechanism} in the following:
\begin{definition}[Quantized Mechanism]\label{Def7}
The \textit{quantized  mechanism}  ${m}^q=(\bs{f}^q,\bs{h}^q)$ is given by
\begin{subequations}\label{quan}
\begin{align}
   {f}_{i}^q(\bs{c})&=f^*_i\left(\bs{Q}(\bs{c})\right),~\forall i\in\mathcal{I},\\
    {h}_i^q(\bs{c})&=c_i\cdot {f}_i^q(\bs{c})+\int_{c_i}^{\bar{c}_i}{f}^q_i(z,\bs{c}_{-i})dz,~\forall i\in\mathcal{I}.\label{quantized}
\end{align}
\end{subequations}

\end{definition}
\subsubsection{Computational complexity} We illustrate an example quantized mechanism in Fig. \ref{illl}. As in Fig. \ref{illl}, the integral in \eqref{quantized} is a Riemann sum, i.e., a computationally efficient finite sum approximation of $\int_{c_i}^{\bar{c}_i}{f}^*_i(z,\bs{c}_{-i})dz$. Specifically, given a quantization step size $\Delta_Q$, computing the Riemann sum in \eqref{quantized} requires one to
compute $f_i^*(c_i,\bs{Q}_{-i}(\bs{c}_{-i}))$ for at most $\ceil{\frac{\bar{c}_i-\underline{c}_i}{\Delta_Q}}$ points
for each source $i$, where $\bs{Q}_{-i}(\bs{c}_{-i})=\{Q_{j}({c}_{j})\}_{j\neq i}$. Therefore, the overall computational overhead is given by $\mathcal{O}(I/\Delta_Q)$. 

\subsection{Properties of the Quantized Mechanism}

In this subsection, we study the properties of the quantized mechanism in \eqref{quan}.
We note that $f_i^q(c_i,\bs{c}_{-i})$ remains non-increasing in $c_i$ for all $i\in\mathcal{I}$. Hence,
based on the characterizations of the IC and  the IR in Theorem \ref{T3} and \eqref{IRC-M}, we have:
\begin{corollary}
The quantized mechanism ${m}^q$ satisfies \textit{IC} and \textit{IR} conditions in  \eqref{IC} and \eqref{IR}.
\end{corollary}


We next study the performance of the quantized mechanism in \eqref{quan} in terms of the destination's overall cost.
To understand how well the quantized mechanism in \eqref{quan} approximates the optimal mechanism, we derive the following lemma:
\begin{lemma}\label{L5}
The aggregate virtual cost function in Definition \ref{D7} is differentiable in $\bs{c}$ and satisfies
\begin{align}
     \frac{\partial \Psi(\bs{c},f_{\rm agg}(\bs{c}))}{\partial c_i}=\phi_i'(c_i)f_i^*(\bs{c})\leq L_{i, \phi}f_{i,\max}, \forall c_i\in\mathcal{C}_i, \forall i\in\mathcal{I},
\end{align}
where $L_{i,\phi}$ is the Lipschitz constant of $\phi_i(c_i)$.
\end{lemma}
Lemma \ref{L5} is a direct application of the envelop theorem in \cite{Envelope}.
When the PDF of the source $i$'s sampling cost $\gamma_i(c_i)$ is differentiable, it follows that 
\begin{align}
    L_{i,\phi}\triangleq\max_{c_i\in\mathcal{C}_i}
    \left[2-\frac{\Gamma_i(c_i)\gamma_i'(c_i)}{\gamma_i^2(c_i)}\right],~\forall c_i\in\mathcal{C}_i,~\forall i\in\mathcal{I}.
\end{align}
Lemma \ref{L5} characterizes an upper bound of the incremental changes of the aggregate virtual cost $\Phi(\bs{c},f_{\rm agg}(\bs{c}))$ in $c_i$, based on which
we can next show that the quantized mechanism is approximately optimal:
\begin{proposition}\label{P2}
The quantized mechanism leads to a bounded quantization loss compared to the optimal mechanism, given by
\begin{align}
J(\bs{f}^q)-J(\bs{f}^*)\leq \sum_{i\in\mathcal{I}}L_{i,\phi}f_{i,\max}\Delta_Q.
\end{align}
\end{proposition}

\arx{We present the proof of Proposition \ref{P2} in Appendix \ref{ProofP2}.}
This shows how the quantization loss depends on the quantization step size $\Delta_Q$. Therefore, the quantized mechanism enables us to make tradeoffs between the quantization loss captured by Proposition \ref{P2} and the computation overhead of  $\mathcal{O}(I/\Delta_Q)$ by tuning $\Delta_Q$.

\subsection{Numerical Studies}
    	\begin{figure}[t]
  	\centering
		\subfigure[Uniform Distribution]{\includegraphics[scale=.3]{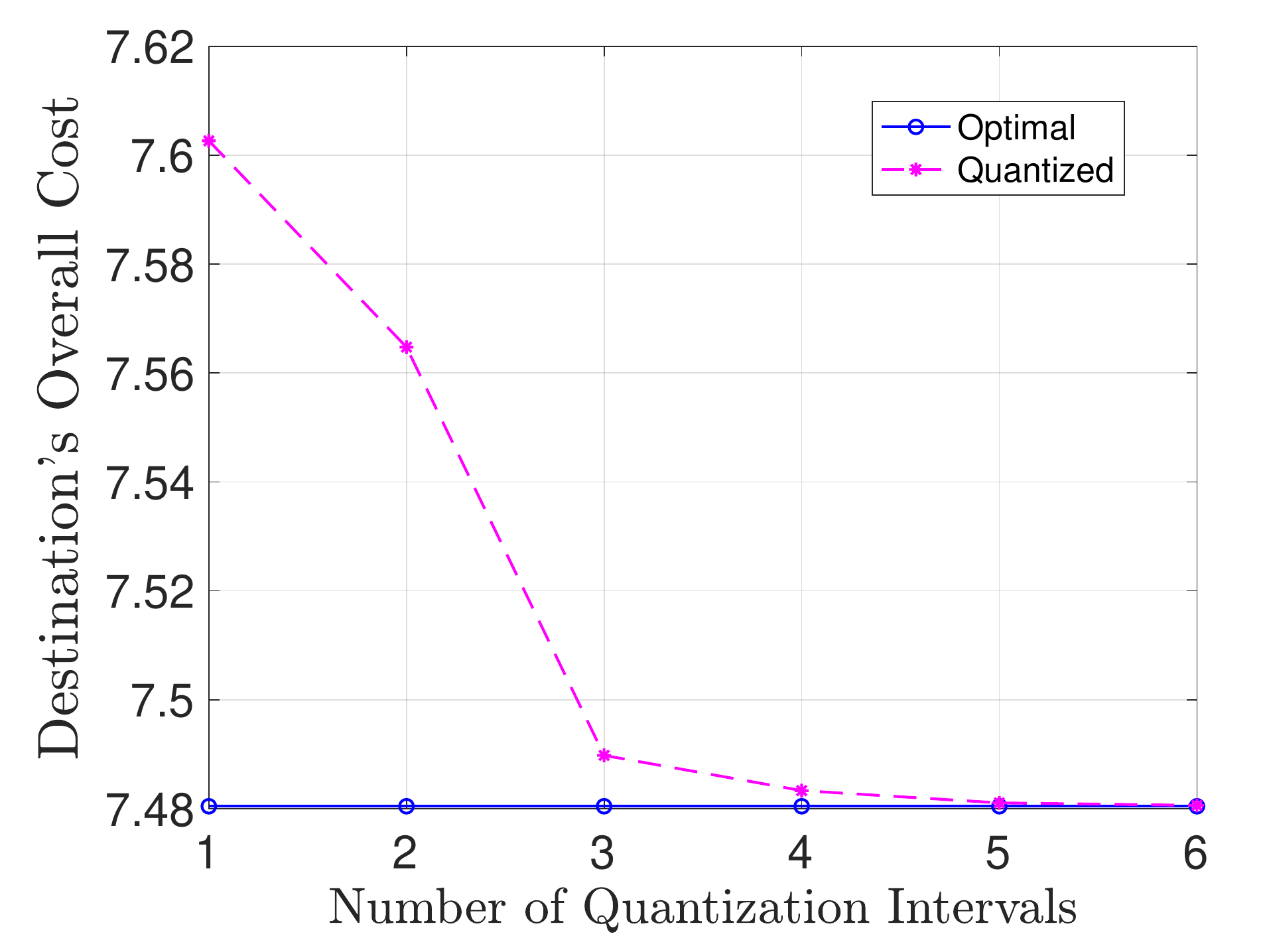}}
		\subfigure[Truncated Exponential Distribution]{\includegraphics[scale=.3]{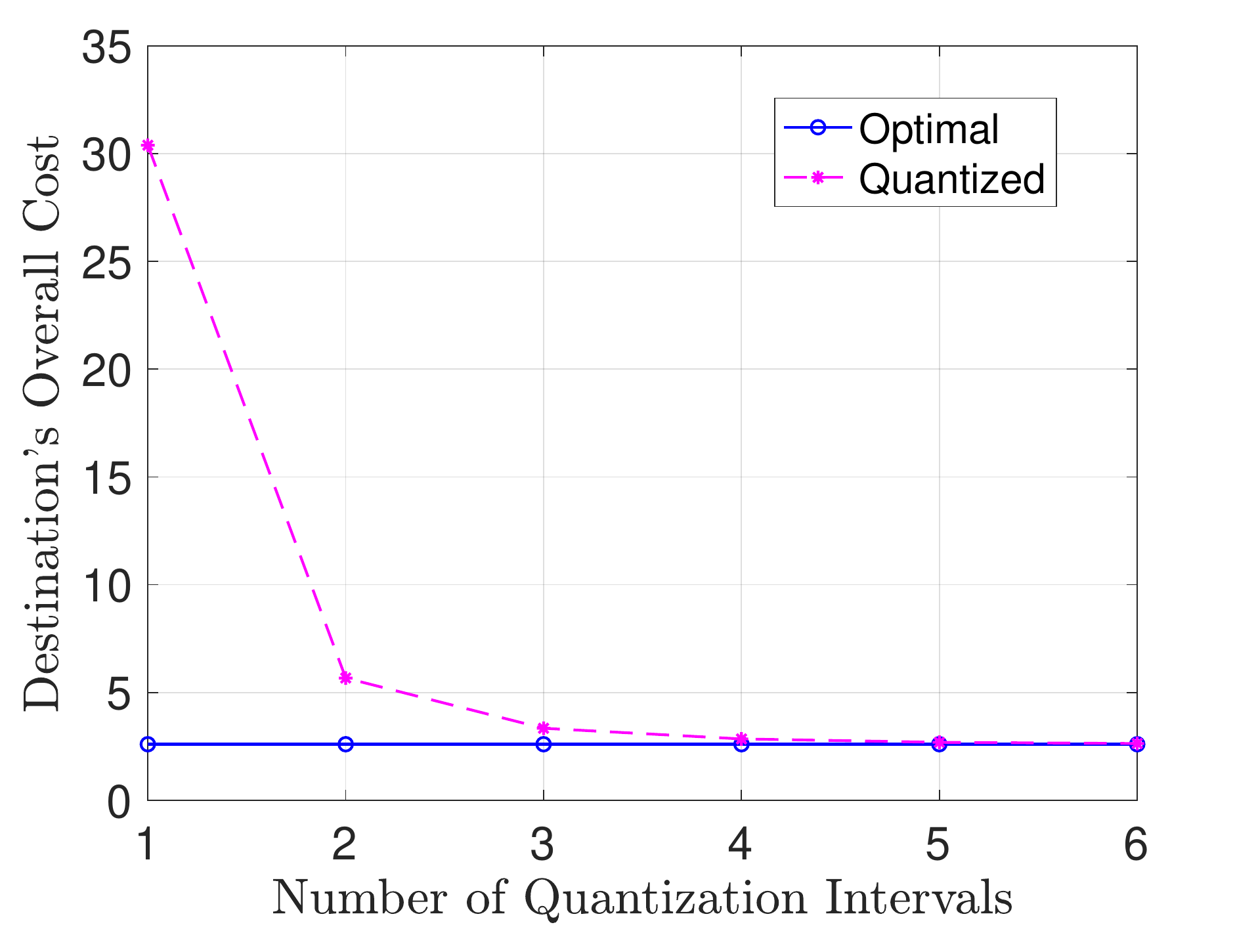}}
		\vspace{-0.05cm}
		\caption{Impact of the number of quantization intervals (i.e., $\ceil{(\bar{c}-\underline{c})/\Delta_Q}$) for a single-source system on
	 the quantization loss (i.e., the difference of the source's overall costs between the quantized mechanism and the optimal mechanism) under different distributions of the source's sampling cost. }\label{Quan}
		\vspace{-0.75cm}
	\end{figure} 

We provide numerical results in Fig. \ref{Quan} to understand the impact of the number of quantization intervals (i.e., $\ceil{(\bar{c}-\underline{c})/\Delta_Q}$) on the quantization loss in a single-source system. 
We consider two classes of distributions of the source's sampling cost, namely the uniform distribution and the truncated exponential distribution.\footnote{These two distributions of costs are also considered in \cite{Cost}.}

\subsubsection{Uniform Distribution}

We first consider the uniform distribution over the interval $[5,30]$, the corresponding Lipschitz constant is $L_{\phi}=2$. Fig. \ref{Quan}(a) shows that, when there are at least three quantization intervals, the quantization only incurs negligible loss (less than $0.1\%$ of the minimal overall cost), which verifies Proposition \ref{P2}. 
\subsubsection{Truncated Exponential Distribution}

We consider the truncated exponential distribution over the interval $[0,30]$ with a PDF:
\vspace{-0.5cm}
\begin{align}
    \gamma(c)=\frac{\exp(- c)}{1-\exp(- \bar{c})}.\label{Eq50}
\end{align}
The corresponding Lipschitz constant is $L_{\phi}=1+\exp(\bar{c})=1+\exp(30)$. Fig. \ref{Quan}(b) shows that,  the quantization only incurs a negligible loss when the number of quantization intervals exceeds a threshold value of around $4$. However, the quantization loss grows rapidly as the number of quantization intervals decreases beyond this point, due to the large $L_{\phi}$. 

The above examples show that the quantized mechanism can achieve approximate optimality with only a moderate number of quantization levels.


\section{General Virtual Cost Function}\label{Gen}
The results in Theorems \ref{T2} and \ref{T4} require non-decreasing virtual cost functions $\phi_i(c_i)$.
In this section, we extend our results to the more general case in which $\phi_i(c_i)$ may not be always non-decreasing. We first introduce the \textit{Ironing} technique \cite{Myerson} and start with the following definitions:

\begin{definition}[Convex Hull]\label{CH}
The convex hull of a function $h(x):\mathcal{X}\rightarrow \mathbb{R}$ is defined as
\begin{align}
    \tilde{h}(x)\triangleq \min\left\{\lambda h(x_1)+(1-\lambda)h(x_2)|\lambda\in [0,1], x_1, x_2\in\mathbb{R}~{\rm and}~\lambda x_1+(1-\lambda) x_2=x \right\}.
\end{align}
\end{definition}

\begin{definition}[Ironed Virtual Cost]\label{IVC}
Define the  cumulative virtual cost as $\Phi_i(c_i)=\int_{\underline{c}_i}^{c_i} \phi_i(t)\gamma_i(t)dt$. Let $\tilde{\Phi}_i(c_i)$ be the convex hull of $\Phi_i(c_i)$. 
 Let ${[a_{i,1}, b_{i,1}], ..., [a_{i,k}, b_{i,k}]}$
be the intervals such that $\tilde{\Phi}_i(c_i)<\Phi_i(c_i)$.
Each source $i$'s ironed virtual cost is
\begin{align}\label{iron}
    \tilde{\phi}_i(c_i)\triangleq\begin{cases}
    \frac{\int_{a_{i,k}}^{b_{i,k}}\phi_i(t)\gamma_i(t)dt}{\Gamma(b_{i,k})-\Gamma(a_{i,k})},&~{\rm if}~c_i\in[a_{i,k},b_{i,k}],\\
    \phi_i(c_i),&~{\rm otherwise}.
    \end{cases}
\end{align}
\end{definition}

\begin{figure}[t]
\centering
	\subfigure[]{\includegraphics[scale=.3]{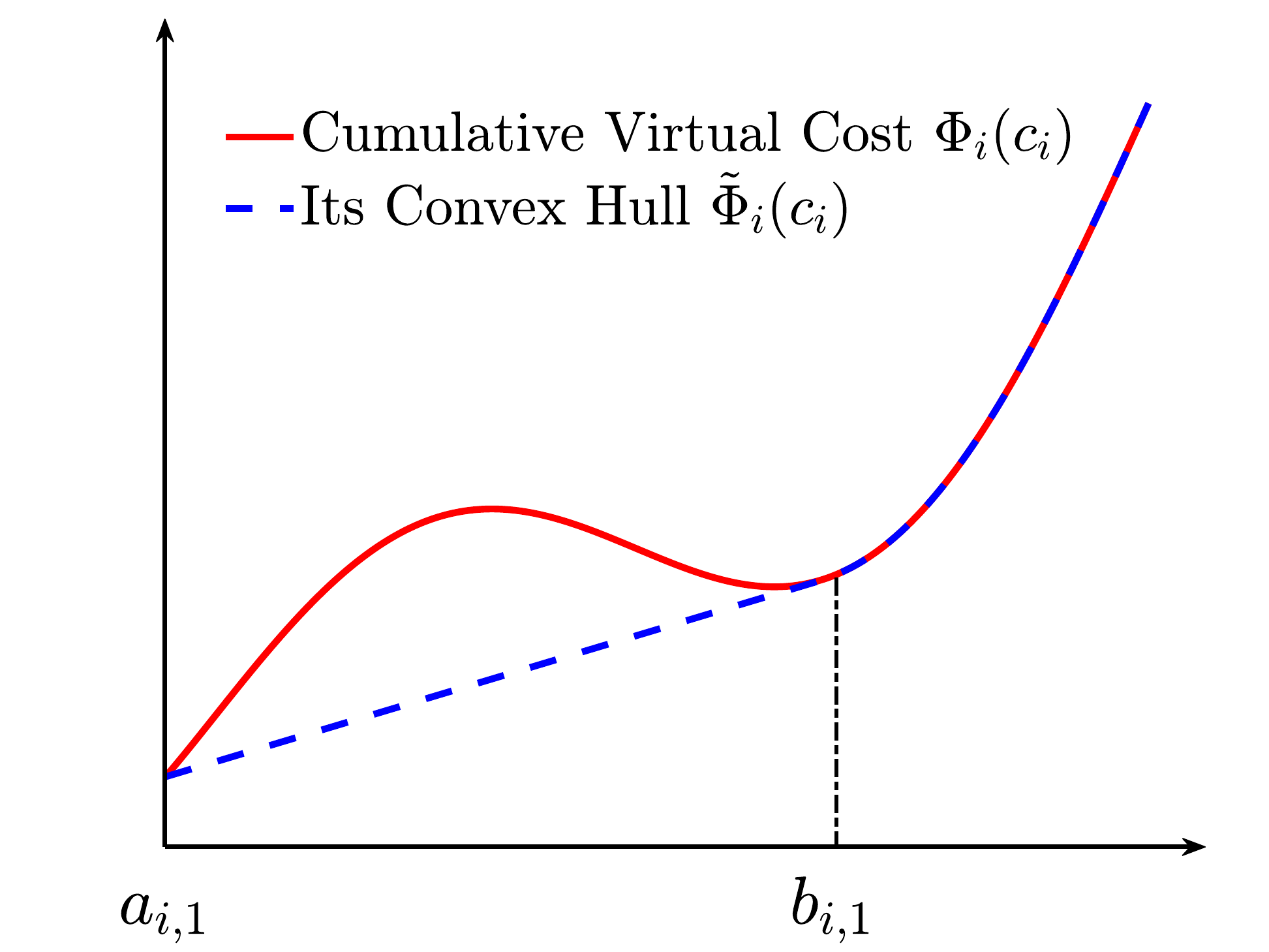}}
		\subfigure[]{\includegraphics[scale=.3]{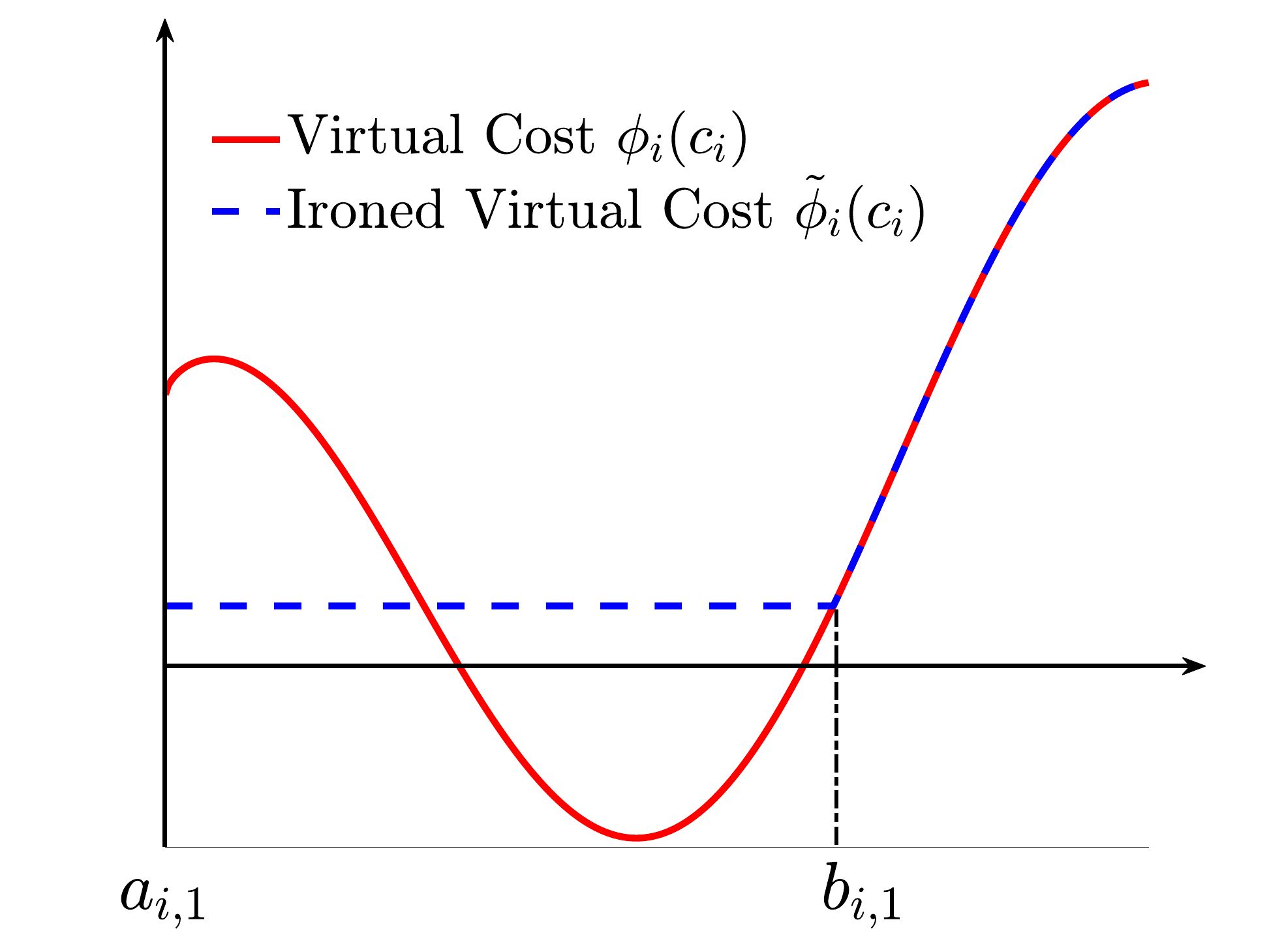}}
		\vspace{-0.2cm}
		\caption{Illustration of the convex hull of the cumulative virtual cost $\tilde{\Phi}_i(c_i)$ (a) and the ironed virtual cost $\tilde{\phi}_i(c_i)$ (b). }\label{ironfig}
		\vspace{-0.75cm}
\end{figure} 

To understand Definitions \ref{CH} and \ref{IVC}, we present an illustrative example in Fig. \ref{ironfig} of the cumulative virtual cost $\Phi_i(c_i)$, its convex hull $\tilde{\Phi}_i(c_i)$, and the ironed virtual cost $\tilde{\phi}_i(c_i)$. In Fig. \ref{ironfig}(a), the convex hull $\tilde{\Phi}_i(c_i)$ straightened out the region $[a_{i,1},b_{i,1}]$. Fig. \ref{ironfig}(b) shows that the ironed virtual cost $\tilde{\phi}_i(c_i)$ is constant over the straightened region $[a_{i,1},b_{i,1}]$, hence the ironed virtual cost $\tilde{\phi}_i(c_i)$ is always non-decreasing in $c_i$. 

We define an ironed version of the aggregate virtual cost:
\begin{definition}[Ironed Aggregate Virtual Cost]
Let $\tilde{\Psi}(\bs{c},f_{\rm agg}(\bs{c}))$ be the aggregate virtual cost function, defined as
\vspace{-0.5cm}
\begin{subequations}
\begin{align}
     \tilde{\Psi}(\bs{c},f_{\rm agg}(\bs{c}))\triangleq\min_{\bs{f}}\quad~& \sum_{i\in\mathcal{I}}\tilde{\phi}_i(c_i) {f}_i(\bs{c})\\
    ~{\rm s.t.} \quad~ & \sum_{i\in\mathcal{I}} {f}_i(\bs{c})=f_{\rm agg}(\bs{c}),\\
    &~f_i(\bs{c})\in [0,f_{i,\max}],~\forall i\in\mathcal{I}.
\end{align}
\end{subequations}
\end{definition}


We next show that replacing ${\Psi}_i(\bs{c})$ by $\tilde{\Psi}_i(\bs{c})$ in \eqref{z4} leads to the following:
\begin{theorem}\label{T5}
For any general virtual cost function $\phi_i(c_i)$, the optimal solution $
\bs{f}^*(\cdot)$ to the problem in \eqref{Problem} satisfies
\begin{align}
    f_{(i)}^*(\bs{c})=\left[f_{\rm agg}^{\rm iron}(\bs{c})-\sum_{j=1}^{i-1}f^*_{(j)}(\bs{c})\right]_0^{f_{(i),\max}}, ~\forall i\in\mathcal{I},  \label{z66}
\end{align}
where $f^{\rm iron}_{\rm agg}(\bs{c})$ satisfies
\begin{align}
g\left(\frac{1}{f^{\rm iron}_{\rm agg}(\bs{c})}\right)\frac{1}{f^{\rm iron}_{\rm agg}(\bs{c})}-G\left(\frac{1}{f^{\rm iron}_{\rm agg}(\bs{c})}\right) \in \partial \tilde{\Psi}(\bs{c}, f^{\rm iron}_{\rm agg}(\bs{c})),~\forall \bs{c}\in\mathcal{C}.
\end{align}
\end{theorem}
Intuitively, since the ironed virtual costs are non-increasing, the resultant $f^*_i(c_i,\bs{c}_{-i})$ in \eqref{z66} is non-decreasing in $c_i$ for all $i\in\mathcal{I}$, which ensures  \eqref{IC} and \eqref{IR} even when relaxing the $f'_i(\bs{c})>0$ constraints in \eqref{Problem-2b}.
\arx{We present the complete proof of Theorem \ref{T5} in Appendix \ref{ProofT4}.}



\section{Performance Comparison}\label{PC} 
In this section, we present analytical and numerical studies to understand when the optimal mechanism in Theorem \ref{T4} and the quantized mechanism in \eqref{quan} are most beneficial, and the impacts of system parameters on the proposed mechanisms.

	
	
\subsection{Benchmarks}

For performance comparison, we introduce a benchmark mechanism and a lower bound achieved by a pricing scheme assuming complete information.
We first define the  benchmark mechanism inspired by the second-price auction \cite{survey1} and will show that such a mechanism satisfies the constraints in \eqref{IC} and \eqref{IR}:
\begin{definition}[Benchmark Mechanism]\label{BM}
The destination only selects source $(1)$ (i.e. the one with the least reported sampling cost)  and 
subsidizes it with the second smallest (reported) sampling cost;
the update policy rule $\bs{f}_B(\tilde{\bs{c}})$ is to minimize the destination's overall cost in \eqref{Overall}, i.e.,
\begin{subequations}\label{Naive}
\begin{align}
    f_{i,B}({\bs{c}}) =&\begin{cases}\arg\min_{f_i\in[0,f_{i,\max}]} \left[f_i G\left(\frac{1}{f_i}\right)-f_i {c}_{(2)}\right], &\quad {\rm if} \quad i=\arg\min {c}_j,\\
0,~&\quad {\rm otherwise}.
\end{cases}\label{xn} \\
    h_{i,B}({\bs{c}})=& f_i({\bs{c}}) {c}_{(2)},~\forall i\in\mathcal{I}, \label{pn}
\end{align}
 \end{subequations}
 where $c_{(2)}$ is the second smallest sampling cost and is set to be $\bar{c}$ when there is only one source.
 \end{definition}
Note that the benchmark mechanism satisfies the IC constraint \eqref{IC} and the IR constraint \eqref{IR}.
 This is because, the source with the smallest sampling cost cannot achieve a higher payoff than truthful reporting,
 as  \eqref{Naive} only depends on the second smallest  report.
 
The following is a lower bound achieved by a pricing scheme assuming complete information:
 \begin{definition}[Complete-Information Pricing Scheme]\label{LB}
Under the complete information setting, the destination subsidizes each source their exact sampling costs; the update policy rule $\bs{f}$ aims to minimize its long-term average AoI cost and the long-term average payments. Mathematically,
\begin{subequations}
\begin{align}
    &\bs{f}_{F}(\bs{c})=\arg\min_{\bs{f}(\cdot)\in\mathcal{F}}\quad \left[G\left(\frac{1}{\sum_{i\in\mathcal{I}}f_i(\bs{c})}\right)\sum_{i\in\mathcal{I}}f_i(\bs{c})+\sum_{i\in\mathcal{I}}c_i\cdot f_i(\bs{c})\right],\\
    &{h}_{i,F}(\bs{c})={f}_{i,F}(\bs{c})c_i,~\forall i\in\mathcal{I}.
\end{align}
\end{subequations}
 \end{definition}
 Such a pricing scheme leads to a lower bound of the destination's overall cost.
Due to the assumption of complete information, this pricing scheme does not satisfy the incentive compatibility constraint \eqref{IC}, while it achieves the individual rationality constraint in \eqref{IR}.

\subsection{Single-Source Systems}

For the single-source systems, as in Section \ref{Single}, we consider both a uniform distribution and a truncated exponential distribution of the sources' sampling costs. We aim to understand when the proposed mechanisms are most beneficial, compared against the benchmark mechanism.

\subsubsection{Uniform Distribution}

We first compare the performance under a uniform distribution of the sampling cost on the interval $[\underline{c},\bar{c}]$; the destination has a power AoI cost 
 $g(x)=x^\alpha$, $\alpha>0$. We assume $f_{\rm \max}$ is sufficiently large.
The benchmark mechanism in Definition \ref{BM} leads to an overall cost of the destination of
\begin{align}
J_{B}=\left[\bar{c}\left(1+\frac{1}{\alpha}\right)\right]^{\frac{\alpha}{1+\alpha}}.\label{benchmark1}
\end{align}
The lower bound  of the overall cost under complete information in Definition \ref{LB} is given by:
\begin{align}
    J_{C}= \frac{\bar{c}^{\frac{1+2\alpha}{1+\alpha}}-\underline{c}^{\frac{1+2\alpha}{1+\alpha}}}{\bar{c}-\underline{c}}
    \frac{1+\alpha}{1+2\alpha}\left(1+\frac{1}{\alpha}\right)^{\frac{\alpha}{1+\alpha}}.
\end{align}
	\begin{figure*}[t]
		\centering
		\subfigure[]{\includegraphics[scale=.27]{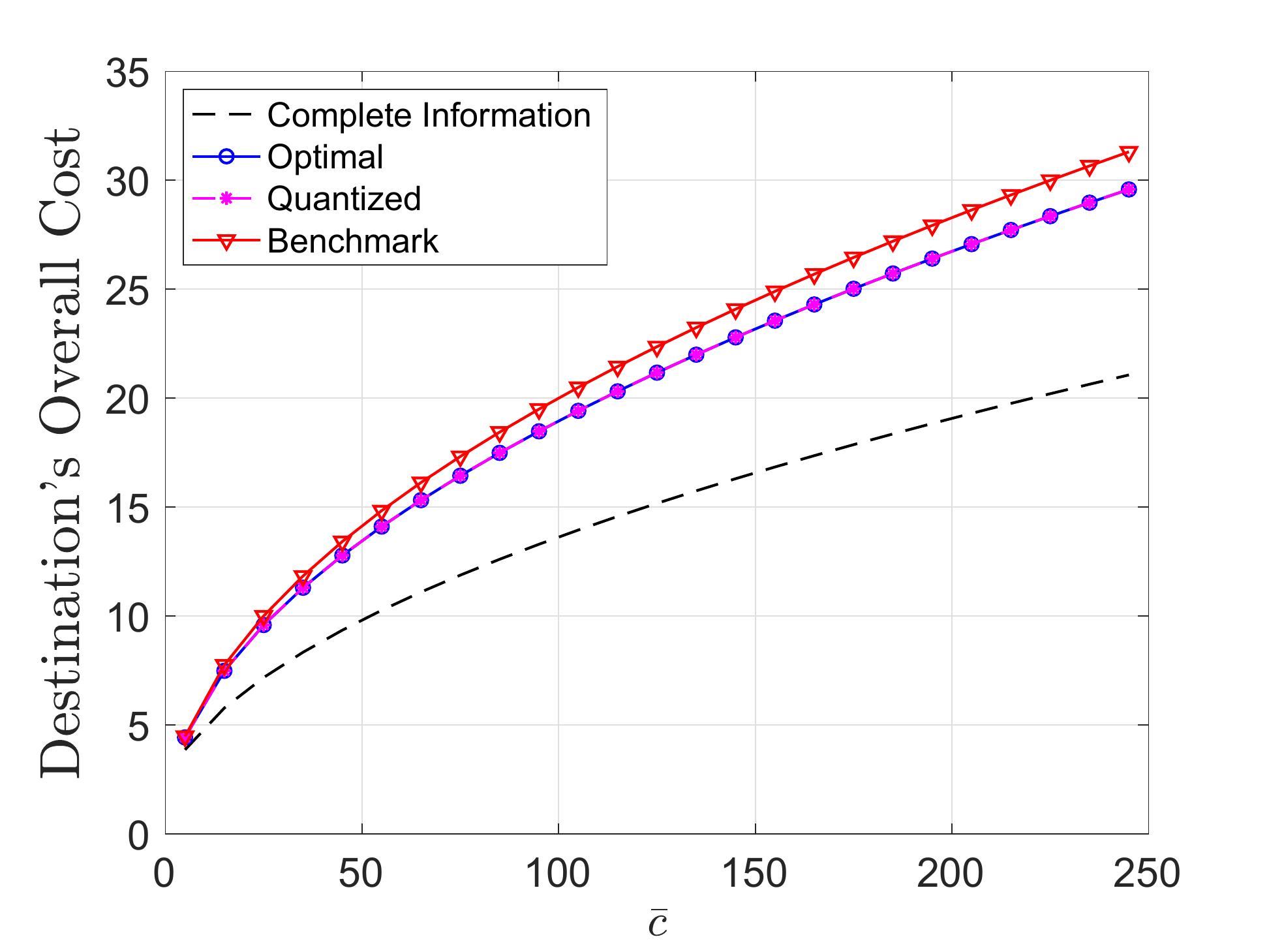}}
		\subfigure[]{\includegraphics[scale=.27]{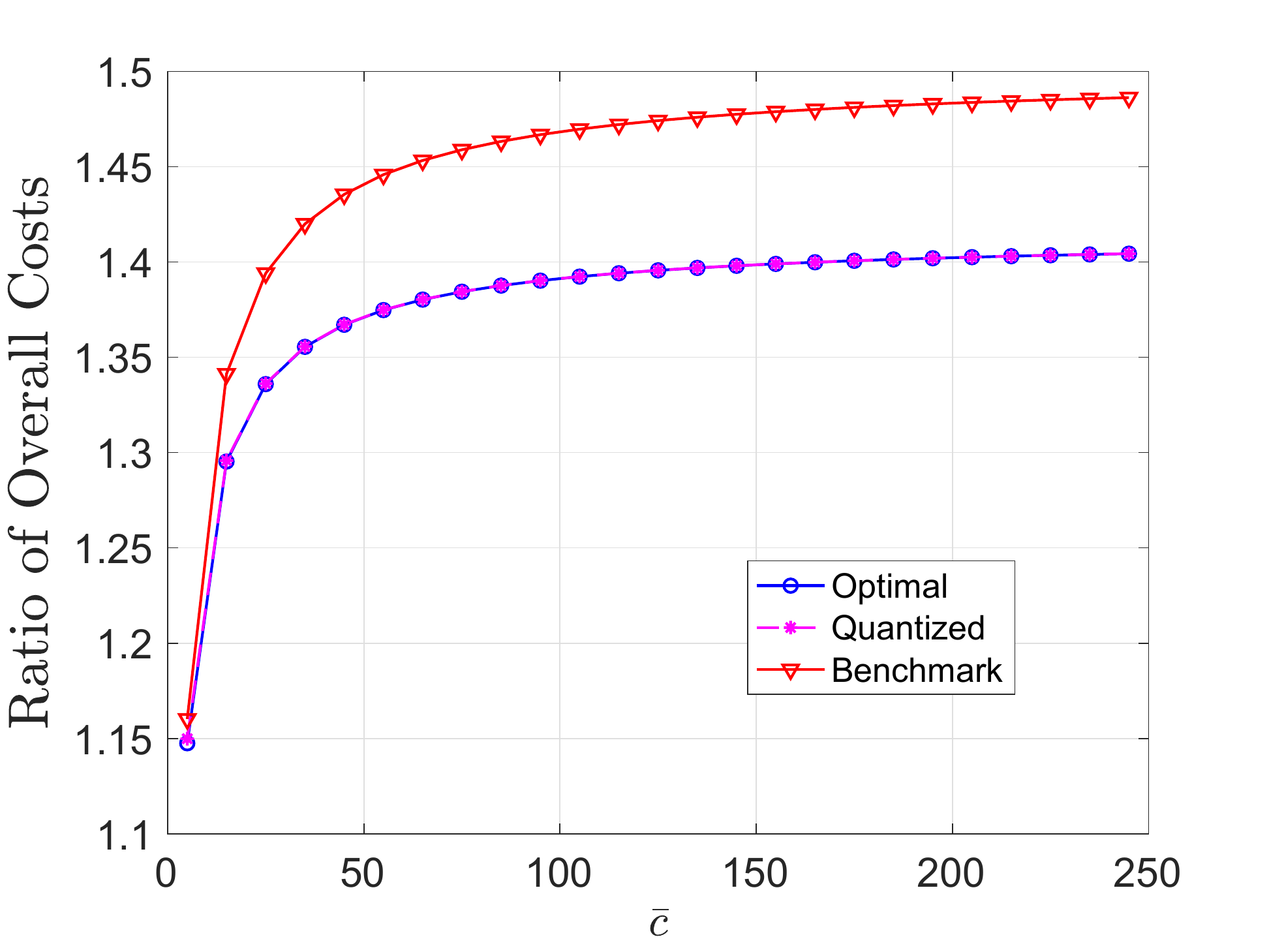}}
		\subfigure[]{\includegraphics[scale=.27]{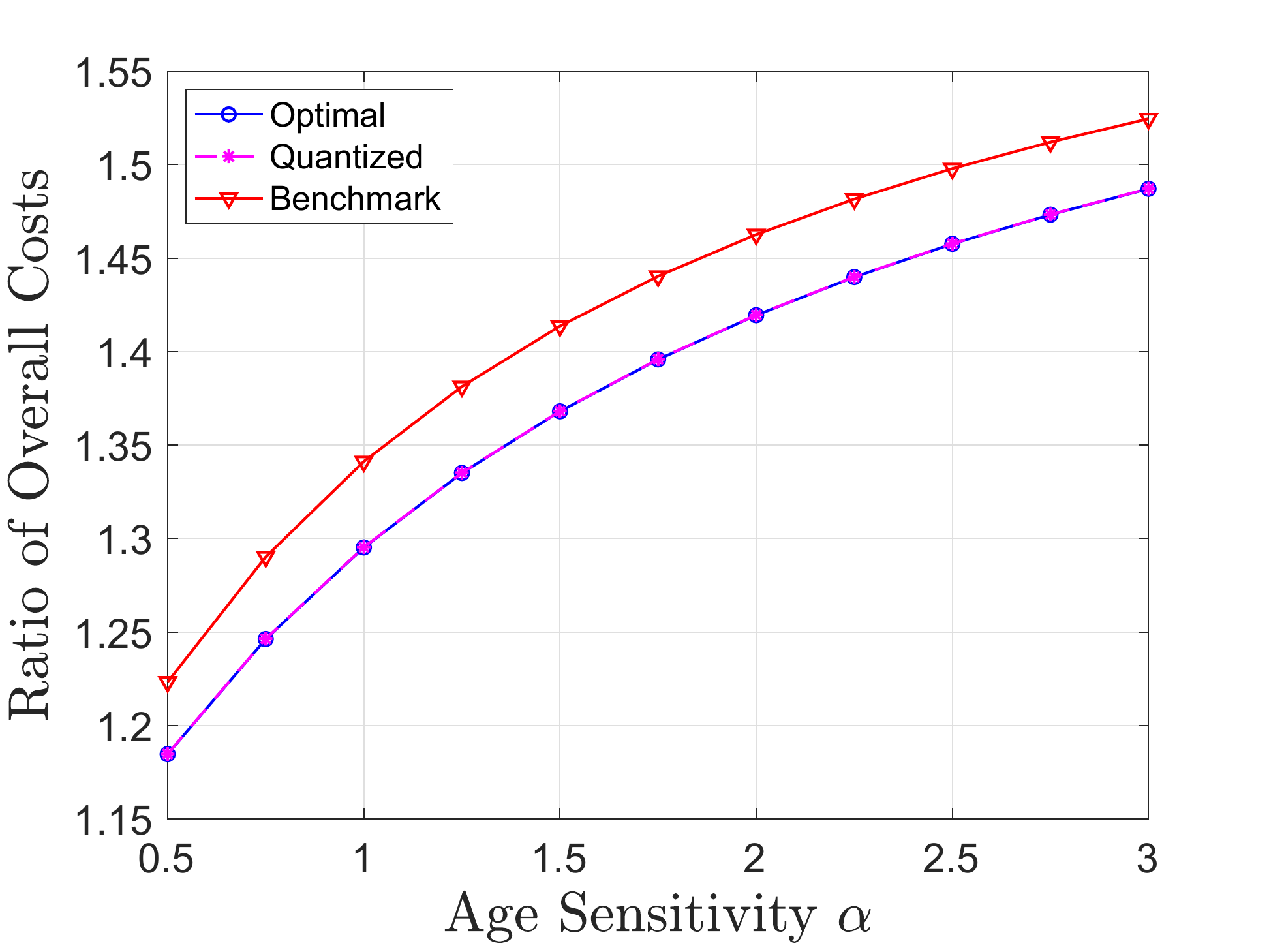}}
		\vspace{-10pt}
		\caption{Performance comparison under the uniform distribution with different upper bounds of the sampling cost $\bar{c}$ in (a), (b) and with different age sensitivity coefficients $\alpha$ in (c). We set the quantization step size to be  $\Delta_Q=1$. We set $\alpha=1$ in (a) and (b) and $\bar{c}=30$ in (c).}\label{fig1}
		\vspace{-25pt}
	\end{figure*}
Hence, we have 
\begin{align}
    \frac{J_B}{J_C}=\left(1+\frac{\alpha}{1+\alpha}\right)\left[\frac{(\bar{c}-\underline{c})\bar{c}^{\frac{\alpha}{1+\alpha}}}{\bar{c}^{\frac{1+2\alpha}{1+\alpha}}-\underline{c}^{\frac{1+2\alpha}{1+\alpha}}}\right]\leq 1+\frac{\alpha}{1+\alpha},\label{Z111}
\end{align}
indicating that, under the uniform distribution, the benchmark mechanism incurs a bounded loss due to private information.

On the other hand, the optimal mechanism in Theorem \ref{T2} leads to an overall cost of
\begin{align}
    J^*=  \left[2\left(1+\frac{1}{\alpha}\right)\right]^{\frac{\alpha}{1+\alpha}}\left(\frac{1+\alpha}{1+2\alpha}\right)\frac{\left\{(c-\frac{\underline{c}}{2})^{\frac{1+2\alpha}{1+\alpha}}\right\}\Big|_{\underline{c}}^{\bar{c}}}{\bar{c}-\underline{c}}.
\end{align}
We can thus obtain the following upper bound:
\begin{align}
    \frac{J^*}{J_C}=\frac{\left\{(c-\frac{\underline{c}}{2})^{\frac{1+2\alpha}{1+\alpha}}\right\}\Big|_{\underline{c}}^{\bar{c}}}{\bar{c}^{\frac{1+2\alpha}{1+\alpha}}-\underline{c}^{\frac{1+2\alpha}{1+\alpha}}}2^{\frac{\alpha}{1+\alpha}}\leq 2^{\frac{\alpha}{1+\alpha}}.\label{Z2}
\end{align}
Equations 
\eqref{Z111} and \eqref{Z2} imply that, under the uniform distribution, the performance gain of the optimal mechanism compared to the benchmark mechanism is limited. 

In Fig.   \ref{fig1}, we numerically compare the performances of the proposed optimal mechanism, the benchmark mechanism, and the complete information lower bound. 
We observe a relatively small gap between the proposed optimal mechanism and the benchmark mechanism under different $\bar{c}$ in Fig. \ref{fig1}(a) and different $\alpha$ in Fig. \ref{fig1}(c).
In Fig. \ref{fig1}(b), both the proposed optimal mechanism and the benchmark mechanism approach their upper bounds in \eqref{Z111} and \eqref{Z2}. 
In addition, the quantized optimal mechanism incurs negligible quantization loss under the uniform distribution.

\subsubsection{Truncated Exponential Distribution}

In this subsection, we consider an exponential distribution of the sampling cost truncated on the interval $[0,\bar{c}]$, i.e., assuming $\underline{c}=0$. The corresponding PDF is given in \eqref{Eq50}.
 Note that the performance of the benchmark mechanism only depends on $\bar{c}$ instead of the specific distribution of $c$. Hence, the overall cost is the same as in \eqref{benchmark1}.
	\begin{figure*}[t]
		\centering
		\subfigure[]{\includegraphics[scale=.27]{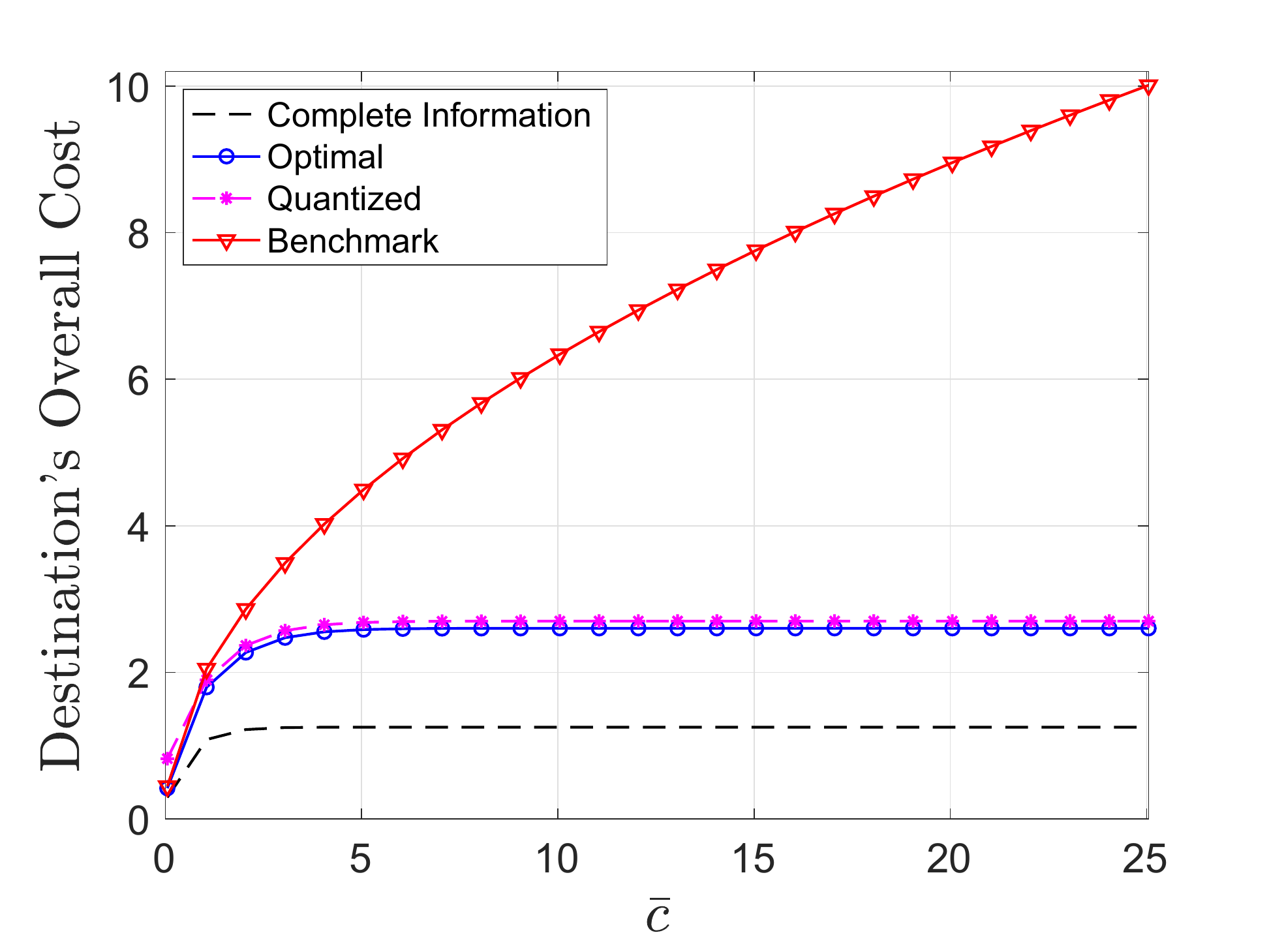}}
		\subfigure[]{\includegraphics[scale=.27]{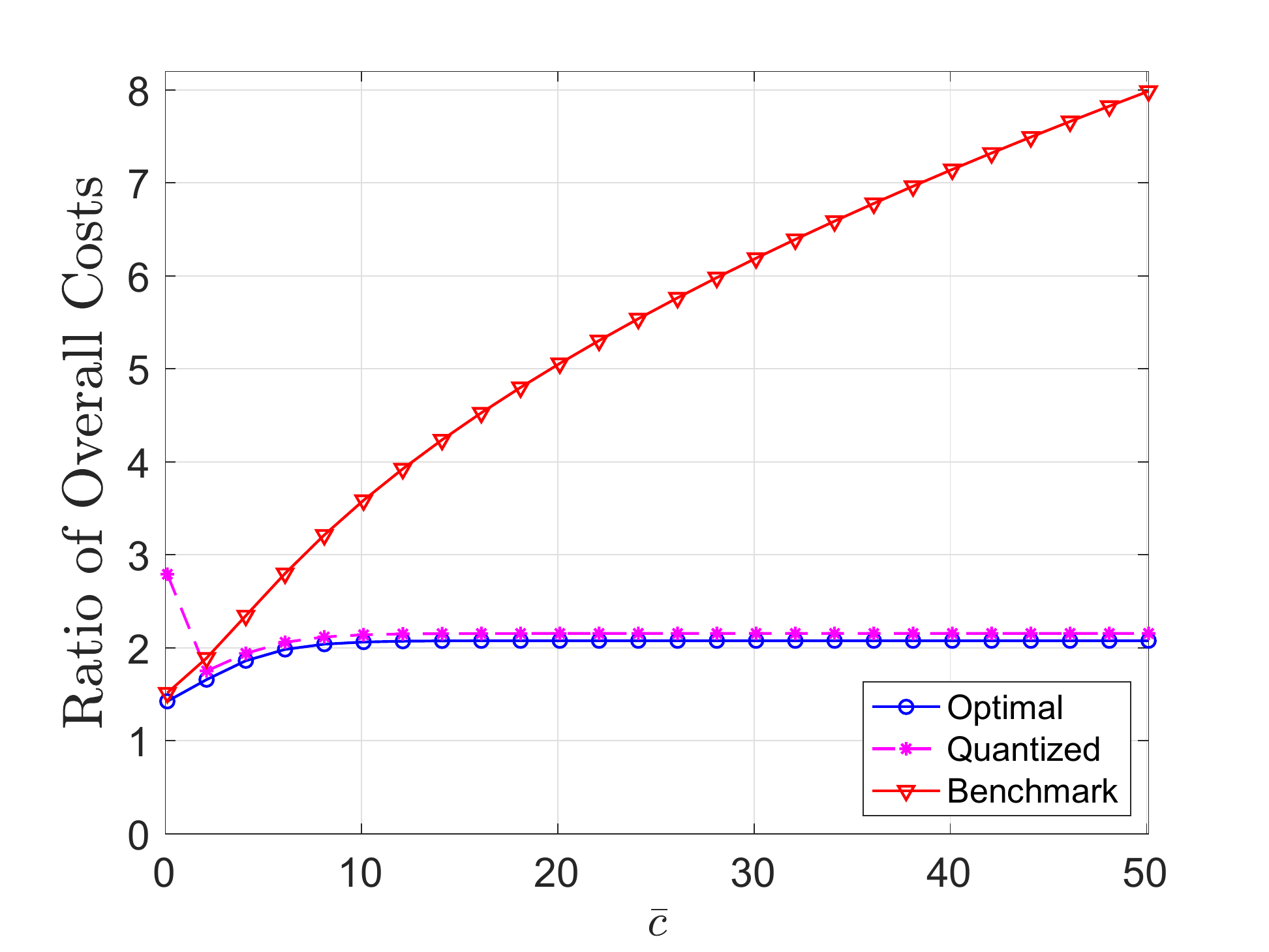}}
		\subfigure[]{\includegraphics[scale=.27]{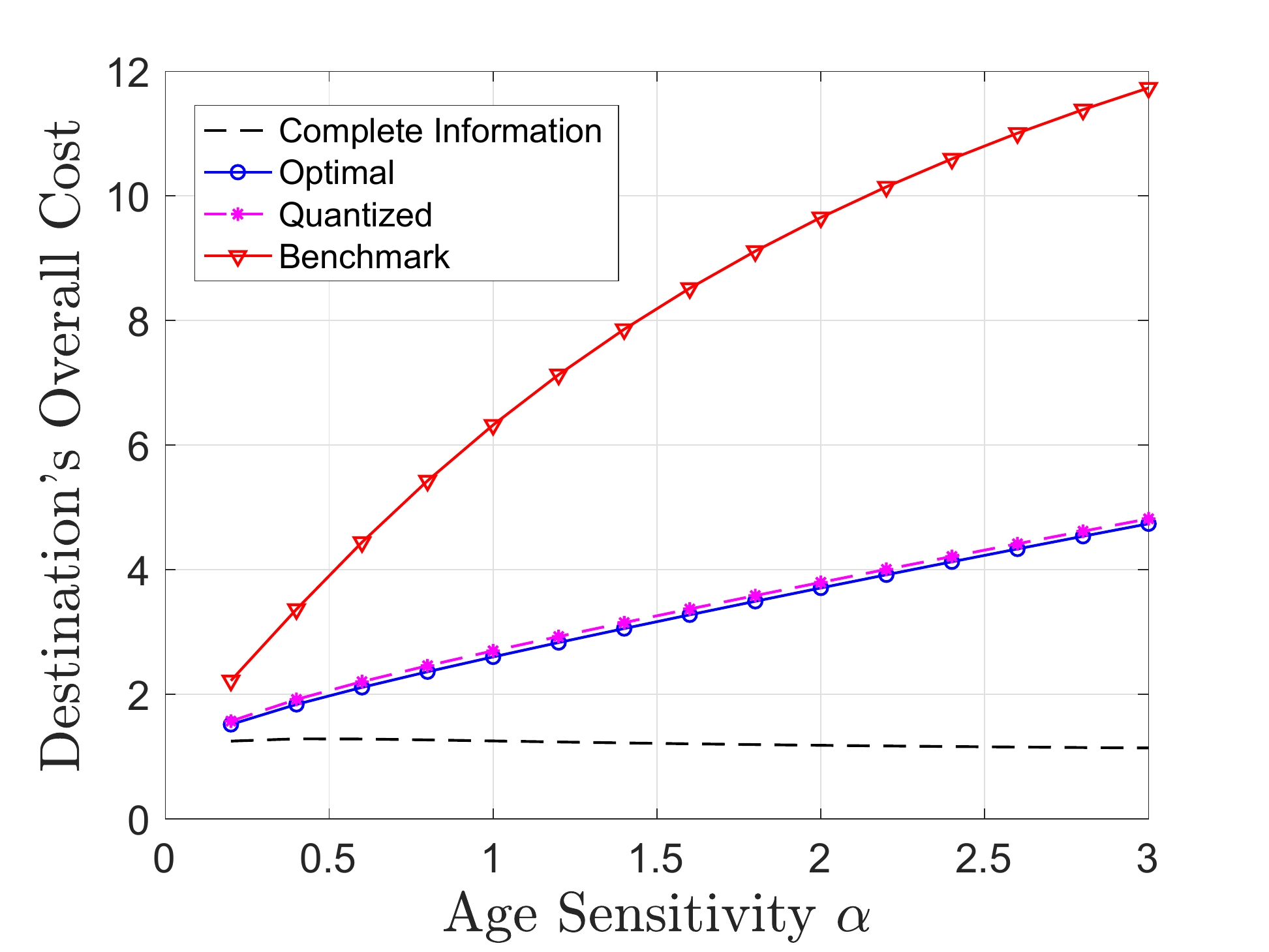}}
		\vspace{-10pt}
		\caption{Performance comparison under the truncated exponential distribution with different upper bounds of the sampling cost $\bar{c}$ in (a), (b) and with different age sensitivity coefficients $\alpha$ in (a), (c). We set the quantization step size to be  $\Delta_Q=1$ and $\underline{c}=0$. In (a) and (b), we set $\alpha=1$. In (c), we set $\bar{c}=30$.}\label{fig2}
		\vspace{-25pt}
	\end{figure*}
 The lower bound of the overall cost under complete information is:
 \begin{align}
      J_{C}=\frac{(1+\frac{1}{\alpha})^{\frac{\alpha}{1+\alpha}}}{1-e^{-\bar{c}}}\left[\Gamma\left(\frac{2\alpha}{1+\alpha}+\frac{1}{\alpha},0\right)-\Gamma\left(\frac{2\alpha+1}{1+\alpha},\bar{c}\right)\right],\label{Jc}
 \end{align}
 where ${\displaystyle \Gamma (s,x)=\int _{x}^{\infty }t^{s-1}\,\exp(-t)\,{\rm {d}}t}$ is the incomplete gamma Function. Note that \eqref{Jc} converges to a finite value when $\bar{c}\rightarrow \infty$.

 The optimal mechanism leads to an overall cost of:
\begin{align}
    J^*=\frac{(1+\frac{1}{\alpha})^{\frac{\alpha}{1+\alpha}}}{1-e^{-\bar{c}}}\int_{0}^{\bar{c}}\left(t-1+\exp(t)\right)^{\frac{a}{1+a}}\exp(-t)dt.
\end{align}
 
Fig. \ref{fig2}(a) shows that overall costs of the destination under the optimal mechanism and the complete information lower bound converge as $\bar{c}$ increases.
 The benchmark mechanism in this case leads to an unbounded overall cost as $\bar{c}$ increases. In Fig. \ref{fig2}(b), we observe relatively small gaps between the optimal mechanism and the complete information lower bound, and between the optimal mechanism and the quantized mechanism. In particular, we have $J^*/J_C\approx2$ when $\bar{c}\geq 10$. Therefore, we have shown that under the truncated exponential distribution, both proposed mechanisms can lead to unbounded benefits, compared against the benchmark mechanism.
 Fig. \ref{fig2}(c) shows that the optimal and the quantized mechanisms become more beneficial when the destination is more sensitive to the AoI, compared with the benchmark mechanism.

\subsection{Multi-Source Systems}

	\begin{figure*}[t]
		\centering
		\subfigure[]{\includegraphics[scale=.27]{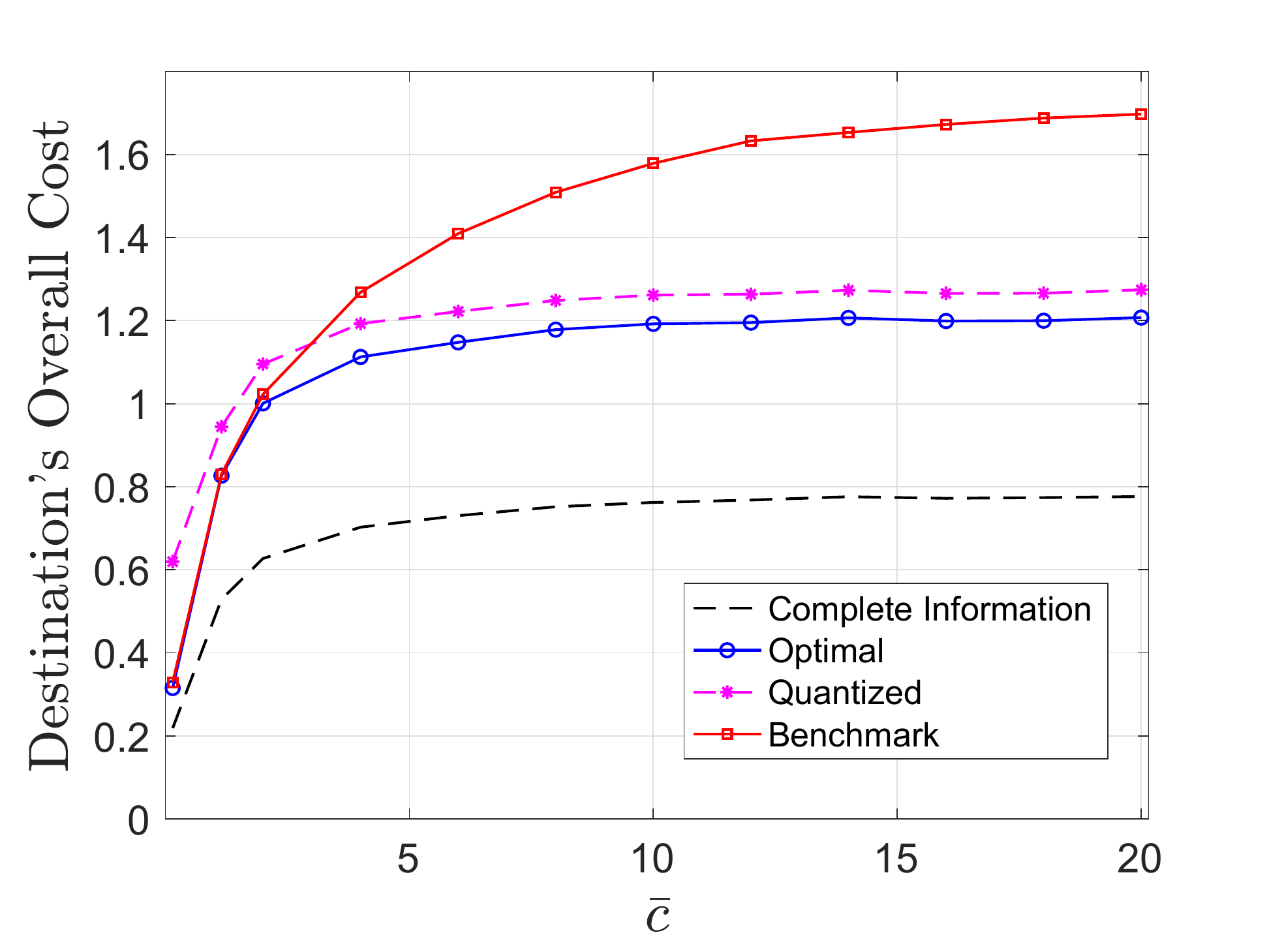}}
		\subfigure[]{\includegraphics[scale=.27]{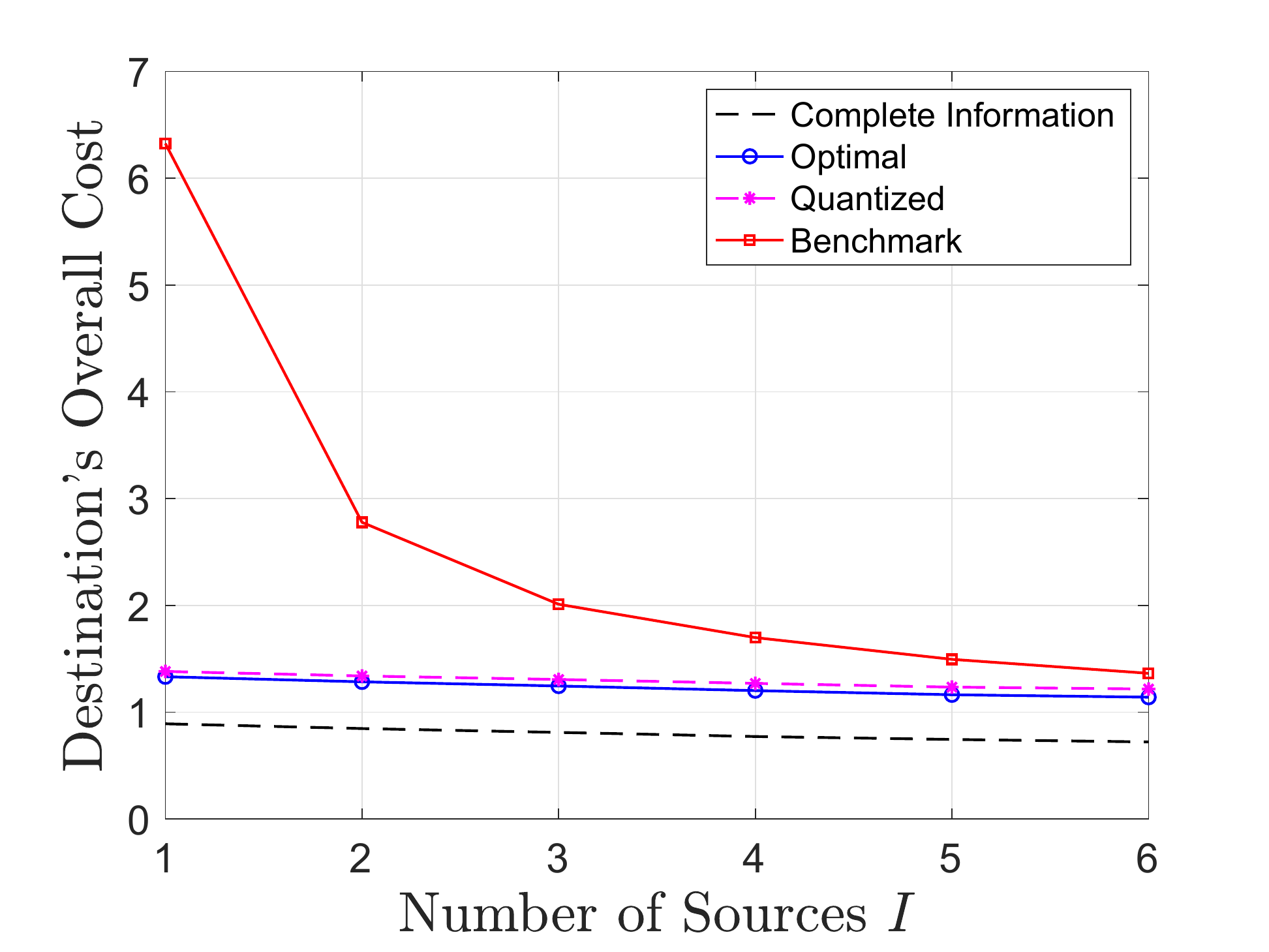}}
		\subfigure[]{\includegraphics[scale=.27]{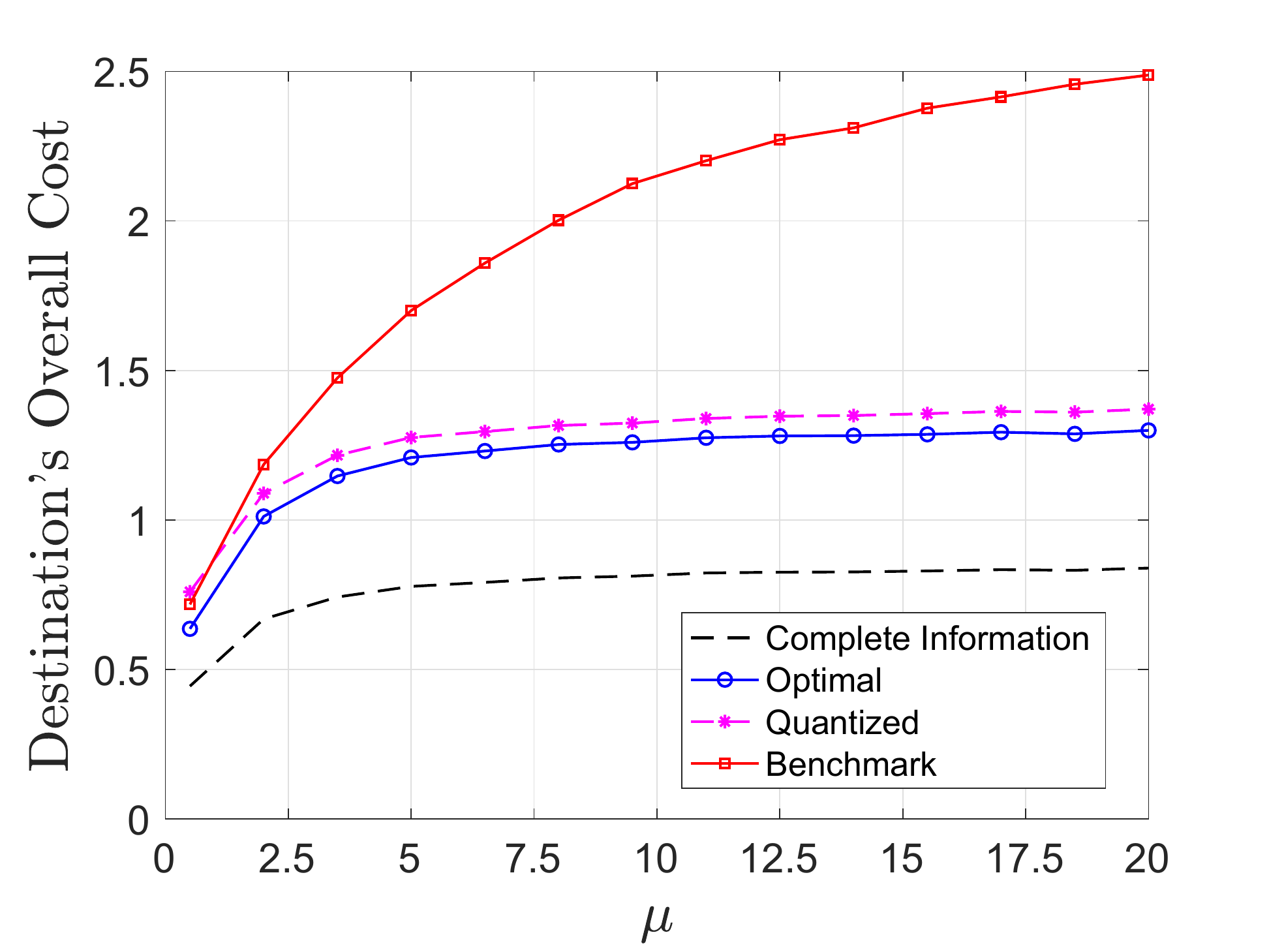}}
		\vspace{-10pt}
		\caption{Performance comparison with different upper bounds of the sampling cost $\bar{c}$ in (a), different number of sources $I$ in (b), and different distribution parameter $\mu$ in (c). We set  $\Delta_Q=0.5$ and consider a linear AoI cost: $f(\Delta_t)=\Delta_t$. In (a) and (b), $\mu_1=0.5$ and $\mu_i=2$ for all $i\neq 1$. In (a) and (c), we set $I=4$. In (b) and (c), we set $\bar{c}=20$. }\label{fig3}
		\vspace{-25pt}
	\end{figure*}
	
We next perform numerical studies to evaluate the impacts of system parameters on the performance for the multi-source systems. 

 The destination has a linear AoI cost:
 $g(x)=x$; $f_{i,\rm \max}$ is sufficiently large for each $i\in\mathcal{I}$.\footnote{Small $f_{i,\rm \max}$ may increase the performance gaps between the benchmark mechanism and the proposed mechanisms, as the benchmark mechanism only assigns one source. }
 We further assume exponential distributions for all sources truncated on the interval $\mathcal{C}_i=\mathcal{C}_j=[0,\bar{c}]$, i.e., the PDF of the sampling cost for each source $i$ is 
    \begin{align}
            \gamma_i(c_i)=
            \frac{\mu_i \exp(-\mu_i c_i)}{1-\exp(-\mu_i \bar{c}_i)},~\forall i\in\mathcal{I}.
    \end{align}

\subsubsection{Impacts of $\bar{c}$}

Fig. \ref{fig3}(a) shows that overall costs of the destination under the optimal mechanism and the complete information lower bound increase as $\bar{c}$ increases. However, the gap between the optimal mechanism and the benchmark is smaller, compared to the gap in Fig. \ref{fig2}(a). Intuitively, the increasing number of sources makes the expected second smallest sampling cost smaller, which
makes incentivizing truthful reporting less costly for the benchmark mechanism and hence makes the performance gap smaller.

\subsubsection{Impacts of the  Number of Sources $I$}

In the second experiment for multi-source systems, we set $\mu_1=0.5$ and $\mu_i=2$ for all $i\neq 1$.
Fig. \ref{fig3}(b) illustrates the impact of the the number of sources $I$ on the performances of the benchmarks and the optimal mechanism. Fig. \ref{fig3}(b) shows that, as $I$ increases, the performances of the complete information lower bound, the optimal mechanism, and the quantized mechanism only slightly decrease, while that of the benchmark mechanism dramatically decreases. 
 As we mentioned, this is because the increasing number of sources reduces the expected second smallest sampling, which
 makes it less costly for the benchmark to induce truthful reports.
 Hence, when there are many sources $I$, the benchmark mechanism may serve as a close-to-optimal solution.

\subsubsection{Impacts of $\mu$}
In our last experiment, 
we set $I=4$, $\mu_1=0.5$, and $\mu_i=\mu$ for all $i\neq 1$ and study the impacts of the parameter $\mu$. 
A larger $\mu$ indicates that the sources other than $i$ have larger expected sampling costs compared to source $i$, i.e., sources are considered more heterogeneous.
Fig. \ref{fig3}(c) shows that, as $\mu$ increases, the performance gaps between the benchmark mechanism and the proposed mechanisms become larger. On the other hand, when $\mu=0.5$, the optimal mechanism only slightly outperforms the benchmark mechanism. 
Therefore, heterogeneity in sources' sampling costs increases the performance gaps between the proposed mechanisms and the benchmark mechanism.

\section{Conclusions}\label{Conclusion}

We have studied the fresh information acquisition problem in the presence of private information. 
We have designed the optimal mechanism to minimize the destination's AoI cost and its payment to the sources, 
while satisfying the truthfulness and individual rationality constraints. We  have further designed a quantized mechanism to tradeoff between optimality and computational complexity.
Our analysis has revealed that the proposed optimal mechanism may lead to an unbounded benefit, compared against a benchmark mechanism,
though this gain depends on the distribution of the sampling cost. Our numerical results have shown that both proposed mechanisms are most beneficial when there are few sources with heterogeneous sampling costs.

There are a few future directions. The first is to design prior-free mechanisms for systems in which the sources and destinations do not have  distributional information. A second direction is to consider real-time systems in which data sources are also requestors as in some practical systems, e.g., Google Waze \cite{Waze} and GasBuddy \cite{GasBuddy}.
The third future direction is to consider a nonuniform quantized mechanism, which can potentially optimize the quantization performance.

\appendices
\section{Discounted Model}\label{DM}

In this section, we consider a discounted model where the objective values of the destination and the sources are discounted over time.
We will show that this may also lead to equal-spacing and flat-rate mechanisms, as the long-time average model does in Lemma \ref{L1}.

We focus on a single-source single-destination system, i.e., $I=1$. A general mechanism has the following form $m=({S}_k(\tilde{c}),{p}_k(\tilde{c}))$, where we define $S_k(\tilde{c})=\sum_{j=1}^{k}x_j(\tilde{c})$ to be time instance of the $k$-th update 
and $x_j(\tilde{c})$ is the $j$-th interarrival time given the source's report $\tilde{c}.$
   The 
   source's (discounted) payoff is
   \begin{align}
      P(\tilde{c},m)= \sum_{k=1}^{\infty}\delta^{S_k(\tilde{c})}(p_k(\tilde{c})-c).
   \end{align}
The destination's (discounted) overall cost is
   \begin{align}
      J(m)= \mathbb{E}_c\left[\sum_{k=1}^\infty \delta^{S_{k-1}(\tilde{c}^*(m))} F_\delta(S_k(\tilde{c}^*(m))-S_{k-1}(\tilde{c}^*(m)))+\sum_{k=1}^{\infty} \delta^{S_k(\tilde{c}^*(m))} p_k(\tilde{c}^*(m))\right],
   \end{align}
   where $F_\delta(x)\triangleq \int_{0}^x \delta^t f(t)dt$.
Let $m'=({S}'_k(\tilde{c}),{p}'_k(\tilde{c}))$ be an arbitrary optimal mechanism satisfying IC and IR as in \eqref{IC-S} and \eqref{IR-S}, based on which we construct a new equal-spacing and the flat-rate mechanism $m^*=({x}^*,{p}^*)$ such that
\begin{align}
    \sum_{k=1}^\infty \delta^{k{x}^*(\tilde{c})}=\sum_{k=1}^{\infty}\delta^{S'_k(\tilde{c})}~~~{\rm and}~~~\sum_{k=1}^\infty \delta^{k{x}^*(\tilde{c})}p^*(\tilde{c})=\sum_{k=1}^{\infty}\delta^{S'_k(\tilde{c})}p'_k(\tilde{c}).\label{Dz}
\end{align}
 From $\sum_{k=1}^\infty \delta^{k{x}^*(\tilde{c})}=\sum_{k=1}^{\infty}\delta^{S'_k(\tilde{c})}$, we see that
 \begin{align}
    \sum_{k=0}^\infty \delta^{k{x}^*(\tilde{c})}c=\sum_{k=0}^{\infty}\delta^{S'_k(\tilde{c})}c. \label{Dz-2}
 \end{align}
Combining \eqref{Dz-2} and \eqref{Dz} leads to
  \begin{align}
      P(\tilde{c},m')=P(\tilde{c},m^*), ~~\forall \tilde{c}\in\mathcal{C}.
  \end{align}
That is, both mechanisms lead to the same payoff for the destination under any report $\tilde{c}$. Hence, since $m'$ satisfies IR and IC, so does $m^*$. Therefore, it suffices to show that
\begin{align}
    \sum_{k=0}^\infty \delta^{k{x}^*(c)} F_\delta({x}^*(c))\leq \sum_{k=1}^\infty \delta^{S'_{k-1}(c)} F_\delta(S'_k(c)-S'_{k-1}(c)), \forall c\in\mathcal{C}.\label{Ineq}
\end{align}
To show \eqref{Ineq}, we consider the following optimization problem:
\begin{subequations}
\begin{align}
    \min_{\{S_k\}}&\quad \sum_{k=1}^{\infty} \delta^{S_k} F_{\delta}(S_k-S_{k-1})\label{z1-1}\\
    {\rm s.t.}&\quad \sum_{k=1}^\infty \delta^{k{x}^*}=\sum_{k=1}^{\infty}\delta^{S_k}. \label{z1-2}
\end{align}
\end{subequations}
 Let $\lambda^*$ be the optimal dual variable corresponding to \eqref{z1-2}. The necessary condition of the optimal solution $\{S_k^o\}$ is
\begin{align}
   \{S_k^o\}\in\arg \min_{\{S_k\}}&\quad \sum_{k=1}^{\infty} \delta^{S_{k-1}} \left(F_{\delta}(S_k-S_{k-1})+\lambda^*\right).
\end{align}
It follows that the minimal objective value of the problem in  \eqref{z1-1}- \eqref{z1-2} satisfies
\begin{align}
    V\triangleq & \min_{\{S_k\}}\quad \sum_{k=1}^{\infty} \delta^{S_{k-1}} \left(F_{\delta}(S_k-S_{k-1})+\lambda^*\right)\nonumber\\
    =&   F_{\delta}(S_1^o)+\min_{\{S_k\}_{k\geq 2}}\quad \left[\delta^{S_{1}^o} (F_{\delta}(S_2-S_{1}^o)+\lambda^*)+\sum_{k=3}^{\infty} \delta^{S_{k-1}} \left(F_{\delta}(S_k-S_{k-1})+\lambda^*\right)\right]\nonumber\\
    \overset{(a)}{=}&F_{\delta}(S_1^o)+\delta^{S_{1}^o}\min_{\{\tilde{S}_k\}}\quad \sum_{k=1}^{\infty} \delta^{\tilde{S}_{k-1}} \left(F_{\delta}(\tilde{S}_k-\tilde{S}_{k-1})+\lambda^*\right). \label{z23}
\end{align}
where (a) is due to the fact that $\{\tilde{S}_k\}$ satisfies $\tilde{S}_0=0$ and $\tilde{S}_k=S_{k+1}-S_1^o$ for all $k\in\mathbb{N}$. That is, from a dynamic programming perspective, \eqref{z23} implies that the problem at $t=S_1^o$ is the same as the problem at $t=0$. Therefore, we have that the optimal solution to \eqref{z1-1}-\eqref{z1-2} corresponds to a \textit{stationary policy} and is given by
\begin{align}
    S_k^o=k{x}^*, \forall k\in\mathbb{N}.
\end{align}
Combining this and \eqref{Dz} proves \eqref{Ineq}.
     		
To conclude, the above analysis reveals that the discounted model may still lead to an optimal mechanism that is equal-spacing and flat-rate, and thus can be designed in a similar way.

		\section{Proof of Lemma \ref{L1}} \label{ProofL1}
We define 
 \begin{align}
     \bar{x}(\tilde{\bs{c}},m)=\limsup_{K\rightarrow \infty} \frac{\sum_{k=1}^K x_k(\tilde{\bs{c}})}{K}, \label{barx}
 \end{align} 
 as the long-term average interarrival time,
 \begin{align}
     \bar{\pi}_i(\tilde{\bs{c}},m)=\liminf_{K\rightarrow \infty} \frac{\sum_{k=1}^K s_{i,k}(\tilde{\bs{c}})}{K}, \label{barpi}
 \end{align}
as the long-term average frequency for source $i$, and
 \begin{align}
     \bar{p}_i(\tilde{\bs{c}},m)=\frac{\lim_{K\rightarrow \infty} \frac{\sum_{k=1}^K s_{i,k}(\tilde{\bs{c}})p_{i,k}(\tilde{\bs{c}})}{K}}{\bar{\pi}_i(\tilde{\bs{c}},m)},\label{barp}
 \end{align}
as the long-term average payment for source $i$.

Let $m^*=(\mathcal{P}^*,\mathcal{X}^*,\mathcal{S}^*)$ be an arbitrary optimal mechanism. Consider another mechanism $m'=(\mathcal{P}',\mathcal{X}',\mathcal{S}')$ such that, for all $i\in\mathcal{I}$,
\begin{align}
    p_{i,k}'(\bs{c})=\bar{p}_i(\bs{c},m^*)~{\rm and}~x_{k}'(\bs{c})=\bar{x}(\bs{c},m^*)~,\forall k\in\mathbb{N},
\end{align}
and $s_{i,k}'(\cdot)$ is generated according to an i.i.d. distribution across $k$ such that
\begin{align}
    {\rm Pr}(s_{i,k}(\tilde{\bs{c}})=1)=\bar{\pi}_i(\tilde{\bs{c}},m^*),~\forall i\in\mathcal{I}, \forall k\in\mathbb{N}.
\end{align}
Hence, the new mechanism $m'$ is equal-spacing and flat-rate, and its scheduling policy is stationary.
It follows that, $\bar{x}(\tilde{\bs{c}},m^*)=\bar{x}(\tilde{\bs{c}},m')$, $\bar{\pi}_i(\tilde{\bs{c}},m^*)=\bar{\pi}_i(\tilde{\bs{c}},m')$, and $\bar{p}_i(\tilde{\bs{c}},m^*)=\bar{p}_i(\tilde{\bs{c}},m')$ for all $i\in\mathcal{I}$. Under the optimal mechanism $m^*$, each source $i$'s expected payoff satisfies, for all $\tilde{c}_i\in\mathcal{C}_i$, all $\bs{c}\in\mathcal{C}$, and all $i\in\mathcal{I}$,
\begin{align}
\mathbb{E}_{\bs{c}_{-i}}\left[P_i\left(\tilde{c}_i,{\bs{c}}_{-i},m^*\right)]\right]=&\mathbb{E}_{\bs{c}_{-i}}\left[\liminf_{K\rightarrow \infty}\frac{\sum_{k=1}^K s^*_{i,k}(\tilde{\bs{c}})(\bar{p}_{i,k}(\tilde{\bs{c}})-c_i)}{\sum_{k=1}^{K}x_k^*(\tilde{\bs{c}})}\right],\nonumber\\
=&\mathbb{E}_{\bs{c}_{-i}}\left[\frac{ \bar{\pi}_{i}(\tilde{\bs{c}},m^*)(\bar{p}_{i}(\tilde{\bs{c}},m^*)-c_i)}{\bar{x}(\tilde{\bs{c}},m^*)}\right],\nonumber\\
=&\mathbb{E}_{\bs{c}_{-i}}\left[\frac{ \bar{\pi}_{i}(\tilde{\bs{c}},m')(\bar{p}_{i}(\tilde{\bs{c}},m')-c_i)}{\bar{x}(\tilde{\bs{c}},m')}\right],\nonumber\\
=&\mathbb{E}_{\bs{c}_{-i}}\left[P_i\left(\tilde{c}_i,{\bs{c}}_{-i},m'\right)]\right].\label{ProofEq72}
\end{align} 
From \eqref{ProofEq72}, we observe that, if $m^*$ is incentive compatible and individually rational, then $m'$ is also incentive compatible and individually rational. Hence, for all $i\in\mathcal{I}$,
\begin{align}
 c_i\in \arg&\max_{\tilde{c}_i\in\mathcal{C}_i}
    ~\mathbb{E}_{\bs{c}_{-i}}[P_i\left(\tilde{c}_i,{\bs{c}}_{-i},m'\right)],\nonumber\\
&\max_{\tilde{c}_i\in\mathcal{C}_i}
    ~\mathbb{E}_{\bs{c}_{-i}}[P_i\left(\tilde{c}_i,{\bs{c}}_{-i},m'\right)]\geq 0.
\end{align} 

Finally, we have that the destination's ex ante long-term overall cost satisfies
\begin{align}
    J(m')
    &=\mathbb{E}_c\left[\frac{G(\bar{x}(c,m'))+\sum_{i\in\mathcal{I}}\bar{\pi}_i(c,m')\bar{p}_i(c,m')}{\bar{x}(c,m')}\right]\nonumber\\
    &=\mathbb{E}_c\left[\frac{G(\bar{x}(c,m'))+\sum_{i\in\mathcal{I}}\bar{\pi}_i(c,m^*)\bar{p}_i(c,m^*)}{\bar{x}(c,m^*)}\right]\nonumber\\
    &\overset{(a)}{\leq} \mathbb{E}_c\left[\frac{\lim_{K\rightarrow \infty}\sum_{k=1}^KG(x_k(c,m^*))/K+\sum_{i\in\mathcal{I}}\bar{\pi}_i(c,m^*)\bar{p}_i(c,m^*)}{\bar{x}(c,m^*)}\right]\nonumber\\
    &=J(m^*),
\end{align}
where $(a)$ is  due to the convexity of $G(\cdot)$ and Jensen's inequality. This completes the proof.
		
		
\section{
Proof of Theorems \ref{T1} and \ref{T3}}		\label{ProofT1}

We prove Theorem \ref{T3} directly, which also proves Theorem \ref{T1} as it is a special case of Theorem \ref{T3}.

When all sources other than $i$ are truthfully reporting, each source $i$'s expected payoff is $\mathbb{E}_{\bs{c}_{-i}}[P_i(\tilde{c}_i,\bs{{c}}_{-i},m)]$. Recall that $\mathbb{E}_{\bs{c}_{-i}}[P_i(\tilde{c}_i,\bs{{c}}_{-i},m)]$ is also dependent on the source $i$'s sampling cost $c_i$ as well.
For presentation simplicity, we replace 
$\mathbb{E}_{\bs{c}_{-i}}[P_i(\tilde{c}_i,\bs{{c}}_{-i},m)]$ by  $P_i(\tilde{c}_i,c_i)$, $\mathbb{E}_{\bs{c}_{-i}}[f_i(\tilde{c}_i,\bs{c}_{-i})]$ by $f_i(\tilde{c}_i)$, and $\mathbb{E}_{\bs{c}_{-i}}[h_i(\tilde{c}_i,\bs{c}_{-i})]$ by $h_i(\tilde{c}_i)$ in this proof.

We first prove the ``if'' direction and then prove the ``only if'' direction.
The source's payoff is
\begin{align}
    P_i(\tilde{c}_i,c_i)=\tilde{c}_i\cdot f_i(\tilde{c}_i)-\int_{\underline{c}_i}^{\tilde{c}_i}f_i(z)dz  +C-f_i(\tilde{c}_i)\cdot c_i,~\forall i\in\mathcal{I}. \label{proofYASD}
\end{align}
Taking the derivative of \eqref{proofYASD} with respect to $\tilde{c}$ yields:
\begin{align}
    \frac{\partial P_i(\tilde{c}_i,c_i)}{\partial \tilde{c}_i}=(c_i- \tilde{c}_i) f'_i(\tilde{c}_i),~\forall i\in\mathcal{I}.\label{hahaha}
\end{align}
Combining \eqref{hahaha} and the fact that $\frac{\partial P_i(\tilde{c}_i,c_i)}{\partial \tilde{c}_i}\leq0$ $(\geq 0)$ for all $\tilde{c}_i\geq c_i$ $(\tilde{c}_i\leq c_i)$ shows that $P_i(\tilde{c}_i,c_i)$ is maximized at $\tilde{c}_i=c_i$. This completes the proof of the ``if'' direction.
		
We next prove the ``only if'' direction.
 To prove the condition 1) is necessary, let $z_1$ and $z_2$ be arbitrary sampling costs satisfying $z_1\leq z_2$. An incentive compatible $m$ requires:
 \begin{align}
     P_i(z_1,z_2)=h_i(z_1)-f_i(z_1)z_2&\leq h_i(z_2)-f_i(z_2)z_2=P_i(z_2,z_2),~\label{z1}\\
     P_i(z_2,z_1)=h_i(z_2)-f_i(z_2)z_1&\leq h_i(z_1)-f_i(z_1)z_1=P_i(z_1,z_1).\label{z2}
 \end{align}
 Combining \eqref{z1} and \eqref{z2}, we have $(f_i(z_1)-f_i(z_2))(z_1-z_2)\leq 0.$ Because $z_1\leq z_2$, we have $f_i(z_1)\geq f_i(z_2)$. Therefore, $f_i(c_i)$ is non-increasing for all $i\in\mathcal{I}$.
 
 To prove condition 2) is necessary, we consider the first order optimality condition of the source's report $\tilde{c}_i^*$:
 \begin{align}
    h'_i(\tilde{c}_i^*)=c_i\cdot f_i'(\tilde{c}_i^*),~\forall i\in\mathcal{I}.\label{Eq75}
 \end{align}
Combining the incentive compatibility of the mechanism  and \eqref{Eq75} yields
 \begin{align}
     h_i'(c_i)=c_i\cdot f_i'(c_i),~\forall c_i\in\mathcal{C}_i, \forall i\in\mathcal{I}.\label{z11}
 \end{align}
 Taking integrals of both sides of \eqref{z11} yields
 \begin{align}
     \int_{\underline{c}_i}^{c_i}h_i'(z)dz&=\int_{\underline{c}_i}^{c_i} z\cdot f_i'(z)dz\nonumber\\
     \Longrightarrow~~~~h_i(c_i)-h_i(\tilde{c}_i)&= c\cdot f_i(c_i)|_{\underline{c}_i}^{c_i}-\int_{\underline{c}_i}^{c_i}f_i(z)dz,~\forall i\in\mathcal{I}.
 \end{align}
This completes the proof.

\section{Proof of Lemma \ref{L2-1} and Lemma \ref{L2}}\label{ProofL2}
We prove Lemma \ref{L2}, which also proves Lemma \ref{L2-1} as it is a special case.

For any source $i\in\mathcal{I}$, we have
\begin{align}
\mathbb{E}_{\bs{c}}\left[c_i\cdot f_i(\bs{c})+\int_{c_i}^{\bar{c}_i}f(z,\bs{c}_{-i})dz\right]=&~\mathbb{E}_{\bs{c}}\left[c_i\cdot f_i(\bs{c})\right]+\int_{\underline{c}_i}^{\bar{c}_i}\left(\int_{t}^{\bar{c}_i} \mathbb{E}_{\bs{c}_{-i}}[f_i(z,\bs{c}_{-i})] dz\right) \gamma_i(t)  dt,\nonumber\\
\overset{(a)}{=}&~\mathbb{E}_{\bs{c}}\left[c_i\cdot f_i(\bs{c})\right]+\int_{\underline{c}_i}^{\bar{c}_i}\left(\int_{\underline{c}_i}^{z}\gamma_i(t) dt\right) \mathbb{E}_{\bs{c}_{-i}}[f_i(z,\bs{c}_{-i})] dz\nonumber\\
=&~\mathbb{E}_{\bs{c}}\left[c_i\cdot f_i(\bs{c})\right]+\int_{\underline{c}_i}^{\bar{c}_i}\frac{\Gamma_i(z)}{\gamma_i(z)}\cdot \mathbb{E}_{\bs{c}_{-i}}[f_i(z,\bs{c}_{-i})] \gamma_i(z) dz\nonumber\\
=&~\mathbb{E}_{\bs{c}}\left[f_i(\bs{c})\phi_i(c_i)\right],
\end{align}
where (a) involves changing the order of integration. This completes the proof.

 \section{Proof of Theorem \ref{T2}}\label{ProofT2}
 
 We commence with the following definition of a convex functional:
 	\begin{definition}[Convex Functional] 
A functional $h: L^2(\Gamma)\rightarrow \mathbb{R}$ is convex if it satisfies that, for any $f_1, f_2 \in L^2(\Gamma)$, the following inequality holds,
\begin{align}
    \lambda h(f_1)+(1-\lambda)h(f_2)\geq h(\lambda f_1+(1-\lambda)f_2),~\forall \lambda\in[0,1].
\end{align}
\end{definition}
We now prove that $J(f)$ in \eqref{virtualcc} is convex.
Since the perspective\footnote{A perspective of a function $f(x)$ is given by $tf(x/t)$ for all $(x,t)$ such that $t>0$ and $x/t$ is in the domain of $f(\cdot)$.} of a convex function is also convex and the fact that $G$ is convex, we have that $G\left(\frac{1}{x}\right)x+x\phi(c) $ is convex in $x$. It follows that$,~\forall \lambda \in[0,1], c\in\mathcal{C},$
	\begin{align}
	    &\lambda\left[G\left(\frac{1}{f_1(c)}\right)f_1(c)+f_1(c)\phi(c)\right]+(1-\lambda)\left[G\left(\frac{1}{f_2(c)}\right)f_2(c)+f_2(c)\phi(c)\right]\nonumber\\
	    \geq &\left[G\left(\frac{1}{\lambda f_1(c)+(1-\lambda)f_2(c)}\right)(\lambda f_1(c)+(1-\lambda)f_2(c))+(\lambda f_1(c)+(1-\lambda)f_2(c))\phi(c) ]\right].
	\end{align}
	Hence, we have, $\forall \lambda \in[0,1]$,
	\begin{align}
	    &\mathbb{E}_c\left[\lambda\left[G\left(\frac{1}{f_1(c)}\right)f_1(c)+f_1(c)\phi(c)\right]\right]+\mathbb{E}_c\left[(1-\lambda)\left[G\left(\frac{1}{f_2(c)}\right)f_2(c)+f_2(c)\phi(c)\right]\right]\nonumber\\
	    \geq &\mathbb{E}_c\left[G\left(\frac{1}{\lambda f_1(c)+(1-\lambda)f_2(c)}\right)(\lambda f_1(c)+(1-\lambda)f_2(c))+(\lambda f_1(c)+(1-\lambda)f_2(c))\phi(c) ]\right],
	\end{align}
 which shows the convexity of $J(f)$.

In the following analysis, we relax the $f'(c)\leq 0$ constraint.
 We define the Lagrangian of the problem \eqref{Problem} as
 \begin{align}
    \mathcal{L}(f,\mu,\lambda)
    =&\mathbb{E}_{{c}}\left[G\left(\frac{1}{f(c)}\right)f({c})+\phi(c)f({c})\right]+\mathbb{E}_{{c}}[f({c})\mu({c})+(f_{\max}-f({c}))\lambda({c})]\label{LS}.
\end{align}

To solve \eqref{Problem}, we introduce 
 the G$\hat{a}$teaux derivative (analog to sub-gradient in finite dimensional space) of a functional $h$ in the direction of $w\in L^2(\Gamma)$ at $f\in L^2(\Gamma)$ \cite{Infinite}:
\begin{align}
    d h(f;w)\triangleq \lim_{\epsilon\rightarrow 0^+}\frac{h(f+\epsilon w) -h(f)}{\epsilon}.
\end{align}
According to \cite{Nonlinear,Infinite}, the sufficient and necessary KKT conditions of optimality are
\begin{subequations}\label{z22}
\begin{align}
    d\mathcal{L}(f;w)&\geq 0,~\forall w_i\in L^2,\label{z111}\\
    \mu(c)f(c)&=0,~\forall c\in\mathcal{C},\label{z111-1}\\
    \lambda({c})(f_{\max}-f({c}))&=0,~\forall {c}\in\mathcal{C},\label{z111-2}\\
    f(c)&\in[0,f_{\max}],~\forall \bs{c}\in\mathcal{C}.\label{z111-3}
\end{align}
\end{subequations}
The G$\hat{a}$teaux derivative of the objective in \eqref{Problem} with respect to $f$ in the direction of $w$ is:
\begin{align}
    d \mathcal{L} (f;w)\!=\!\mathbb{E}_c\!\left[w(c)\!\left(\!G\left(\frac{1}{f(c)}\right)- g\left(\frac{1}{f(c)}\right)\!\frac{1}{f(c)} +\phi(c)\!\right)\right]\!+\mathbb{E}_{{c}}[w({c})(\mu({c})-\lambda({c}))].\label{Z1}
\end{align}

Consider the Dirac delta function $\delta(x)\in L^2(\Gamma)$ ($\int_{-\infty}^{\infty}\delta(x)dx=1$ and $\delta(x)=0$ for all $x\neq 0$).
Substituting $w(x)=\delta(x-c)$ for each $c\in\mathcal{C}$ into \eqref{Z1}, it follows that the condition in 
\begin{align}
G\left(\frac{1}{f(c)}\right)- g\left(\frac{1}{f(c)}\right)\frac{1}{f(c)} +\phi(c)+\mu({c})-\lambda({c})=0,~\forall c\in\mathcal{C}.\label{condi}
\end{align}
Combining the conditions in \eqref{condi} and \eqref{z111-1}-\eqref{z111-3} yields \eqref{z3} and \eqref{z3-b}. This completes the proof.


\section{Proof of Lemma \ref{L4}}\label{ProofL4}
After relaxing the $f_i'(\bs{c})\leq 0$ constraints and introducing the new decision $f_{\rm agg}(\cdot)$, the problem in \eqref{Problem-2b} becomes
\begin{align}
    &\min_{\bs{f}(\cdot), f_{\rm agg}(\cdot)}\quad J(\bs{f})\triangleq\mathbb{E}_{\bs{c}}\left[G\left(\frac{1}{\sum_{i\in\mathcal{I}}f_i(\bs{c})}\right)\sum_{i\in\mathcal{I}}f_i(\bs{c})+\sum_{i\in\mathcal{I}}\phi_i(c_i) f_i(\bs{c})\right]\\
    &~~{\rm s.t.}\quad \bs{f}(\cdot)\in\mathcal{F}\triangleq\left\{\bs{f}(\cdot): f_i(\bs{c})\in [0,f_{i,\max}], f_i'(\bs{c})\leq 0,~\forall \bs{c}\in\mathcal{C},~i\in\mathcal{I}\right\}\\
    &~~~~~~~~~ \sum_{i\in\mathcal{I}} f_i(\bs{c})=f_{\rm agg}(\bs{c}),~\forall i\in\mathcal{I}.
\end{align}
which is equivalent to Lemma \ref{L4}. This completes the proof.

\section{Proof of Theorem \ref{T4}}\label{ProofT4}

First,  $\Phi(\bs{c},{f}_{\rm agg}(\bs{c}))$ is convex in ${f}_{\rm agg}(\bs{c})$ as partial minimization preserves convexity. Therefore, a subgradient $g$ of $\Phi(\bs{c},x)$ at $x$ satisfies
\begin{align}
    f(z)\geq f(x)+g \cdot (z-x),~\forall z\in\left[0,\sum_{i\in\mathcal{I}}f_{i,\max}\right].
\end{align}
Hence, we have
the subgradient of $\Phi(\bs{c},{f}_{\rm agg}(\bs{c}))$ is given by
\begin{align}\label{z232}
    \partial_{{f}_{\rm agg}(\bs{c})} \Phi(\bs{c},{f}_{\rm agg}(\bs{c}))=\begin{cases}
    \phi_{(i)}(c_{(i)}),~{\rm if}~f_{\rm agg}\in\left(\sum_{(i)\in\mathcal{I}}f_{(i),\max},\sum_{(i+1)\in\mathcal{I}}f_{(i+1),\max}\right) &\forall (i)\in\mathcal{I},\\
    [\phi_{(i)}(c_{(i)}),\phi_{(i+1)}(c_{(i+1)})],~{\rm if}~f_{\rm agg}=\sum_{(i)\in\mathcal{I}}f_{(i),\max}, &\forall (i)\in\mathcal{I},
    \end{cases}
\end{align}

We define the Lagrangian to be
\begin{align}
    \mathcal{L}(\bs{f},\bs{\mu},\bs{\lambda})
    =&\mathbb{E}_{\bs{c}}\left[G\left(\frac{1}{\sum_{i\in\mathcal{I}}f_i(\bs{c})}\right)\sum_{i\in\mathcal{I}}f_i(\bs{c})+\sum_{i\in\mathcal{I}}\phi_i(c_i)f_i(\bs{c})\right]\nonumber\\
    &+\mathbb{E}_{\bs{c}}[f_i(\bs{c})\mu_i(\bs{c})+(f_{i,\max}-f_i(\bs{c}))\lambda_i(\bs{c})].\label{L}
\end{align}
The G$\hat{a}$teaux derivative of $\mathcal{L}$ w.r.t. $f$ in the direction of $w$ is:
\begin{align}
    d \mathcal{L} (f_i;w_i)\!=&\!\mathbb{E}_c\!\left[w_i(\bs{c})\!\left(\!G\left(\frac{1}{\sum_{i}f_i(\bs{c})}\right)- g\left(\frac{1}{\sum_{i\in\mathcal{I}}f_i(\bs{c})}\right)\!\frac{1}{\sum_{i\in\mathcal{I}}f_i(\bs{c})} +\phi_i(\bs{c})\!\right)\right]\nonumber\\
    &+\!\mathbb{E}_c\!\left[w_i(\bs{c})(\mu_i(\bs{c})-\lambda_i(\bs{c}))\right].
\end{align}
According to \cite{Nonlinear,Infinite}, the sufficient and necessary KKT conditions of optimality are
\begin{subequations}\label{z22}
\begin{align}
    d\mathcal{L}(f_i;w_i)&\geq 0,~\forall w_i\in L^2,\label{z222}\\
    \mu_i(\bs{c})f_i(\bs{c})&=0,~\forall \bs{c}\in\mathcal{C},~i\in\mathcal{I},\label{z222-1}\\
    \lambda_i(\bs{c})(f_{i,\max}-f_i(\bs{c}))&=0,~\forall \bs{c}\in\mathcal{C},~i\in\mathcal{I},\label{z222-2}\\
    f_i(\bs{c})&\in[0,f_{i,\max}],~\forall \bs{c}\in\mathcal{C},~i\in\mathcal{I}.
\end{align}
\end{subequations}

From \eqref{z222}, we have
\begin{align}
    \left(\!G\left(\frac{1}{\sum_{i}f_i(\bs{c})}\right)- g\left(\frac{1}{\sum_{i\in\mathcal{I}}f_i(\bs{c})}\right)\!\frac{1}{\sum_{i\in\mathcal{I}}f_i(\bs{c})} +\phi_i(\bs{c})\!\right)+\mu_i(\bs{c})-\lambda_i(\bs{c})=0,~\forall \bs{c}\in\mathcal{C}, i\in\mathcal{I},
\end{align}
which yields
\begin{align}
    \phi_i(\bs{c})+\mu_i(\bs{c})-\lambda_i(\bs{c})=  \phi_j(\bs{c})+\mu_j(\bs{c})-\lambda_j(\bs{c}),~\forall \bs{c}\in\mathcal{C},  i,j\in\mathcal{I}.
\end{align}
Hence, if $f_i^*(\bs{c})>0$, we must have $\lambda^*_j(\bs{c})>0$ for all $j$ such that $\phi_j(\bs{c})<\phi_i(\bs{c})$. That is, the least expensive (in terms of virtual costs) sources must be fully utilized before expensive sources are utilized.
Therefore, 
\begin{align}
    f_{(i)}^*(\bs{c})=\left[f_{\rm agg}(\bs{c})-\sum_{j=1}^{i-1}f^*_{(j)}(\bs{c})\right]_0^{f_{(i),\max}}, \forall \bs{c}\in\mathcal{C}, \forall i\in\mathcal{I}, \label{y32}
\end{align}
for some $f_{\rm agg}(\cdot)$.
Combining \eqref{z22}  and   $\partial_{{f}_{\rm agg}(\bs{c})} \Phi(\bs{c},{f}_{\rm agg}(\bs{c}))$ in \eqref{z232}, we see that $f_{\rm agg}(\cdot)$ in \eqref{y32} must satisfy \eqref{z4}, which leads to
\eqref{T4-eq}. Finally, by Theorem \ref{T3}, we have proved Theorem \ref{T4}.
 \section{Proof of Proposition \ref{P2}}\label{ProofP2}
 
Define the minimal value of the subproblem  of  the problem in \eqref{Problem-2b} for each $\bs{c}\in\mathcal{C}$,
$$\zeta(\bs{c})\triangleq \min_{f(\bs{c})\in\prod_{i\in\mathcal{I}}[0,f_{i,\max}]}\left[G\left(\frac{1}{\sum_{i\in\mathcal{I}}f_i(\bs{c})}\right)\sum_{i\in\mathcal{I}}f_i(\bs{c})+\sum_{i\in\mathcal{I}}\phi_i(c_i)\cdot f_i(\bs{c})\right].$$
Hence, 
\begin{align}
 \zeta(\bs{c}_a)-\zeta(\bs{c}_b)=\int_{\bs{c}_b}^{\bs{c}_a} \nabla_{\bs{c}}\zeta(\bs{c})\bs{d}\bs{c},~\forall \bs{c}_a, \bs{c}_b\in\mathcal{C}.
\end{align}
It follows that 
\begin{align}
    \zeta(\bs{Q}(\bs{c}))-\zeta(\bs{c})=\int_{\bs{c}}^{\bs{Q}(\bs{c})}\sum_{i\in\mathcal{I}}\phi_i'(c_i)f_i^*(c_i,\bs{c}_{-i})\bs{d}\bs{c}\leq \sum_{i\in\mathcal{I}}\frac{\Delta_Q}{2} f_{i,\max}L_{i,\phi}.
\end{align}
 Hence, we have
 \begin{align}
J(\bs{f}^q)-J(\bs{f}^*)&=
     \mathbb{E}_{\bs{c}}[ \zeta(\bs{c}^q)-\zeta(\bs{c})]+ \mathbb{E}_{\bs{c}}\left[\left(\sum_{i\in\mathcal{I}}\phi_i(c_i)-\phi_i(c_i^q)\right)f_i^*(\bs{c}^q)\right]\nonumber\\
     &\leq \sum_{i\in\mathcal{I}} \frac{\Delta_Q}{2} f_{i,\max}L_{i,\phi}+\sum_{i\in\mathcal{I}} \frac{\Delta_Q}{2} f_{i,\max}L_{i,\phi}=\sum_{i\in\mathcal{I}} {\Delta_Q}f_{i,\max}L_{i,\phi}.
 \end{align}
 \section{Proof of Theorem \ref{T5}}\label{ProofT5}

The destination's overall cost can be rewritten as
\begin{align}
    J(\bs{f})
    =&\int_{\mathcal{C}} \left[G\left(\frac{1}{\sum_{i\in\mathcal{I}}f_i(\bs{c})}\right)\sum_{i\in\mathcal{I}}f_i(\bs{c})+\sum_{i\in\mathcal{I}}f_i(\bs{c})\phi_i(c_i)\right]d \Gamma(\bs{c})\nonumber\\
    =&\int_{\mathcal{C}} \left[G\left(\frac{1}{\sum_{i\in\mathcal{I}}f_i(\bs{c})}\right)\sum_{i\in\mathcal{I}}f_i(\bs{c})+\sum_{i\in\mathcal{I}}f_i(\bs{c})\tilde{\phi}_i(c_i)\right]d \Gamma(\bs{c})\nonumber\\
    &+\int_{\mathcal{C}} \left[\sum_{i\in\mathcal{I}}f_i(\bs{c})[{\phi}_i(c_i)-\tilde{\phi}_i(c_i)]\right]d \Gamma(\bs{c}).\label{iron1}
\end{align}
Let $\bs{f}^\star(\cdot)$ be the minimizer of $$\bs{f}^\star(\cdot)=\arg\min_{\bs{f}(\cdot)\in\mathcal{F}}\mathbb{E}_c\left[G\left(\sum_{i\in\mathcal{I}}f_i(\bs{c})\right)\sum_{i\in\mathcal{I}}f_i(\bs{c})+\sum_{i\in\mathcal{I}}f_i(\bs{c})\tilde{\phi}_i(c_i)\right].$$ 
It follows that
\begin{align}
  &\mathbb{E}_c\left[G\left(\frac{1}{\sum_{i\in\mathcal{I}}f^\star_i(\bs{c})}\right)\sum_{i\in\mathcal{I}}f^\star_i(\bs{c})+\sum_{i\in\mathcal{I}}f^\star_i(\bs{c})\tilde{\phi}_i(c_i)\right] \nonumber\\
  \geq ~&\mathbb{E}_c\left[G\left(\frac{1}{\sum_{i\in\mathcal{I}}f_i(\bs{c})}\right)\sum_{i\in\mathcal{I}}f_i(\bs{c})+\sum_{i\in\mathcal{I}}f_i(\bs{c})\tilde{\phi}_i(c_i)\right],~\forall \bs{f}(\cdot)\in\mathcal{F}.\label{iron2}
\end{align}

Since $\tilde{\Phi}_i(\cdot)$ is the convex hull of $\Phi_i(\cdot)$, if $\Phi_i(c_i)>\tilde{\Phi}_i(c_i)$ in $(a_k,b_k)$ then $\tilde{\phi}'_i(c_i)=0$. It follows that $f^\star_i(\bs{c})$ is constant at $(a_k,b_k)$. Therefore, we have
\begin{align}
    &\int_{a_{k_i}}^{b_{k_i}} \left[ G\left(\frac{1}{\sum_{i\in\mathcal{I}}f_i^\star(\bs{c})}\right)f_i^\star(\bs{c})+f_i^\star(c){\phi}(c)-G\left(\frac{1}{\sum_{i\in\mathcal{I}}f_i^\star(\bs{c})}\right)f_i^\star(\bs{c})+\sum_{i\in\mathcal{I}}f_i^\star(\bs{c})\tilde{\phi}_i(c_i) \right]\gamma(c_i)dc_i
    \nonumber\\
    =&\sum_{i\in\mathcal{I}}\int_{\mathcal{C}_{-i}}\int_{a_{k_i}}^{b_{k_i}}  f_i^\star(\bs{c})\left[{\phi}_i(c_i)-\tilde{\phi}_i(c_i) \right]\gamma_i(c_i)\bs{\gamma}_{-i}(\bs{c}_{-i})d\bs{c}\overset{(a)}{=}0,\label{iron3}
\end{align}
where $(a)$ follows directly from the definition of $\tilde{\phi}_i(\cdot)$ in \eqref{iron}.

Combining \eqref{iron1}, \eqref{iron2}, and \eqref{iron3}, we have that 
$J(f)\leq J(f^\star)$ for every $f$. Therefore, $f^\star$ is the optimal solution to the problem in \eqref{Problem}. By Theorem \ref{T4}, we have proved Theorem \ref{T5}.

		
		
		
		


\begin{thebibliography}{99}

			\bibitem{AoI2}
		S. Kaul, R. D. Yates, and M. Gruteser, ``Real-time status: How often should one update?'' in \textit{Proc. IEEE INFOCOM}, 2012.
		
		\bibitem{AoIEcon3}
B. Li, and J. Liu, ``Can we achieve fresh information with selfish users in mobile crowd-learning?'' in \textit{Proc. WiOpt}, 2019. 






		
		
		

\bibitem{Waze}
Waze Mobile App. [Online]. Available: \url{https://www.waze.com/}

\bibitem{GasBuddy}
GasBuddy Mobile App. [Online]. Available: \url{https://www.gasbuddy.com/}

		\bibitem{AoIEcon1}
		X. Wang and L. Duan, ``Dynamic pricing for controlling age of information,'' in \textit{
Proc. IEEE ISIT,} 2019.
		
		\bibitem{AoIEcon2}
		S. Hao and L. Duan, ``Regulating competition in age of information under network externalities,'' vol. 38, no. 4, pp. 697-710, \textit{IEEE J. Sel. Area Comm.}, 2020.


\bibitem{AoIEcon4}
M. Zhang, A. Arafa, J. Huang, and H. V. Poor, ``Pricing fresh data,'' available online: 2006.16805.

		
		\bibitem{IA2}
		Y. Chen, N. Immorlica, B. Lucier, V. Syrgkanis, and J. Ziani,  ``Optimal data acquisition for statistical estimation,'' in \textit{Proc. ACM EC}, 2018.
		
		\bibitem{IA3}
		J. Abernethy, Y. Chen, C.-J. Ho, and B. Waggoner.  ``Low-cost learning via active data procurement,'' in \textit{Proc. ACM EC}, 2015.
		
		\bibitem{IA4}
		A. Roth and G. Schoenebeck,  ``Conducting truthful surveys, cheaply,'' in \textit{Proc. ACM EC}, 2012.
		
		\bibitem{IA5}
		Y. Cai, C. Daskalakis, and C. H. Papadimitriou.  ``Optimum statistical estimation with strategic data sources,'' in \textit{Proc. COLT}, 2015.
		
		\bibitem{DM1}
		M. Babaioff, R. Kleinberg, and R. Paes Leme. ``Optimal mechanisms for selling information.'' in \textit{Proc. ACM EC},  2012.
		
		
		        \bibitem{IV}
        B. Waggoner, and Y. Chen,  ``Output agreement mechanisms and common knowledge,'' in \textit{AAAI HCOMP}, 2014.
        
		
		
		
		
        







%
%
%
		
		
		

		
		

		
		

		
		
		

		
		

		
		
		\bibitem{AoI4}
		Q. He, D. Yuan, and A. Ephremides, ``Optimal link scheduling for age minimization in wireless systems''. \textit{IEEE Trans. Inf. Theory}, vol. 64, no. 7, pp. 5381-5394, July 2018. 
		
		
		
		
		\bibitem{AoI7}
		Y. Sun, E. Uysal-Biyikoglu, R. D. Yates, C. E. Koksal, and N. B. Shroff, ``Update or wait: How to keep your data fresh,'' \textit{IEEE Trans. Inf. Theory}, vol. 63, no. 11, pp. 7492-7508, Nov. 2017.
		
		
		\bibitem{New1}
		R. Talak, S. Karaman, and E. Modiano.  ``Optimizing information freshness in wireless networks under general interference constraints,'' in \textit{Proc. ACM Mobihoc}, 2018.
		
		\bibitem{New2}
		I. Kadota, A. Sinha, and E. Modiano.  ``Optimizing age of information in wireless networks with throughput constraints,'' in \textit{Proc. IEEE INFOCOM}, 2018.
		


		
 		\bibitem{energy1}
 		X. Wu, J. Yang, and J. Wu, ``Optimal status update for age of information minimization with an energy harvesting source,'' \textit{IEEE Trans. on Green Commun. Netw.}, vol. 2, no.1, pp. 193–204, March 2018.
		
		\bibitem{AoI12}
		A. Arafa, J. Yang, S. Ulukus, and H. V. Poor, ``Age-minimal transmission for energy harvesting sensors with finite batteries: Online policies'', \textit{IEEE Trans. Inf. Theory}, vol. 66, no. 1, pp. 534-556, Jan. 2020.
		
 		\bibitem{energy2}
 		B. T. Bacinoglu, Y. Sun, E. Uysal-Biyikoglu, and V. Mutlu. ``Achieving the age-energy tradeoff with a finite-battery energy harvesting source,'' in \textit{Proc. IEEE ISIT}, 2018. 
		
		
 		\bibitem{NewAge1}
 		B. Zhou and W. Saad, ``Joint status sampling and updating for minimizing age of information in the Internet of Things," \textit{ IEEE Trans. Commun.}, 2019.
		
		\bibitem{NewAge2}
 		Ahmed M. Bedewy, Y. Sun, R. Singh, and N. B. Shroff, ``Optimizing information freshness using low-power status updates via sleep-wake scheduling,” in \textit{Proc. ACM MobiHoc}, 2020.
		
		\bibitem{NewAge3}
		 Tasmeen Zaman Ornee and Y. Sun, ``Sampling for Remote Estimation through Queues:  Age of Information and Beyond,'' submitted to \textit{IEEE/ACM Trans. Netw.}, 2020.
		
		
		
		
 		\bibitem{AoI10}
 		G. D. Nguyen, S. Kompella, C. Kam, J. E. Wieselthier, and A. Ephremides, ``Information freshness over an interference channel: A game theoretic view,"  in \textit{Proc. IEEE INFOCOM}, 2018.
		
		





		\bibitem{Myerson}
		R. B. Myerson,  ``Optimal auction design.'' \textit{Mathematics of Operations Research}, vol. 6, no. 1, pp. 58-73, Feb. 1981.
		
		
		

		\bibitem{mechanism2}
		A. M. Manelli, and D. R. Vincent,  ``Optimal procurement mechanisms,'' \textit{Econometrica}, 1995.
		
		
		\bibitem{mechanism3}
 		F. Naegelen, ``Implementing optimal procurement auctions with exogenous quality,'' \textit{Review of Economic Design}, 7(2), pp.135-153, 2002.
	   
	    
 	   \bibitem{mechanism4}
 	   J. Asker, and E. Cantillon,  ``Procurement when price and quality matter.'' \textit{The Rand journal of economics,}  2010.
		
 	   \bibitem{mechanism5}
 	   R. Burguet, J. J. Ganuza, and E. Hauk,  ``Limited liability and mechanism design in procurement,'' \textit{Games and Economic Behavior,} 2012.
		
 		\bibitem{mechanism6}
 		M. Cary, A. D. Flaxman, J. D. Hartline, and A. R. Karlin,   ``Auctions for structured procurement,'' in \textit{Proc. SODA}, 2008.
		
		\bibitem{Online1}
S. Shalev-Shwartz, ``Online learning and online convex optimization,'' \textit{Foundations and Trends in Machine Learning,} vol. 4, no. 2, pp. 107– 194, 2012.

\bibitem{Online2}
X. He, J. Pan, O. Jin, T. Xu, B. Liu, T. Xu, Y. Shi, A. Atallah, R. Herbrich, S. Bowers, and J. Q. N. Candela, ``Practical lessons from predicting clicks on Ads at Facebook,'' in \textit{Proceedings of the Eighth International Workshop on Data Mining for Online Advertising}, 2014.

		\bibitem{GameTheory}
		R. B. Myerson, \textit{Game theory.} Harvard university press, 2013. 
		
		\bibitem{survey1}
		T. Roughgarden, ``Approximately optimal mechanism design: motivation, examples, and lessons learned.'' \textit{ACM SIGecom Exch.} 2015.
		
		\bibitem{survey2}
	    J. D. Hartline,  ``Mechanism design and approximation,'' Book draft, 2013.
		
		 		\bibitem{mechanism1}
 		I. Segal,  ``Optimal pricing mechanisms with unknown demand,'' \textit{American Economic Review}, 2003.
		
		
		
		\bibitem{approx1}
		D. Lehmann, L. I. O’Callaghan, and Y. Shoham, ``Truth revelation in approximately efficient combinatorial auctions,'' \textit{Journal of the ACM}, vol. 49, no. 5, pp. 577–602, 2002.
		
		\bibitem{approx2}
		N. Nisan and A. Ronen. ``Algorithmic mechanism design.'' \textit{Games and Economic Behavior}, vol. 35, no. 1/2, pp. 166– 196, 2001.
		
		\bibitem{approx3}
		N. Nisan, 2015. ``Algorithmic mechanism design: Through the lens of multiunit auctions.'' \textit{Handbook of Game Theory with Economic Applications},  vol. 4, pp. 477-515, Elsevier, 2015.
		
		\bibitem{approx4}
		N. Nisan and I. Segal. ``The communication requirements of efficient allocations and supporting prices.'' \textit{Journal of Economic Theory}, vol. 129 no. 1, pp. 192–224, 2006.
		
		\bibitem{approx5}
		H. Ge and R. A. Berry.  ``Quantized VCG Mechanisms for polymatroid environments,'' in \textit{Proc.  ACM Mobihoc}, 2019.
		
		\bibitem{approx6}
		H. Ge and R. A. Berry, ``Dominant strategy allocation of divisible network resources with limited information exchange,'' in \textit{Proc. IEEE INFOCOM}, 2018.
		
		 \bibitem{Infinite} 
 D. Butnariu and A. N. Iusem, ``Totally convex functions for fixed points computation and infinite dimensional optimization.'' Norwell, Kluwer Academic Publisher, 2000.

 \bibitem{Nonlinear}
 D. P. Bertsekas, Nonlinear Programming, 2nd ed. Belmont, MA: Athena Scientific, 1999.
		
\bibitem{Cost}
J. Wang, J. Tang, D. Yang, E. Wang and G. Xue, ``Quality-aware and fine-grained incentive mechanisms for mobile crowdsensing,'' in \textit{Proc. IEEE ICDCS}, 2016.


 
\bibitem{Envelope}
P. Milgrom, and I. Segal,  ``Envelope theorems for arbitrary choice sets,'' \textit{Econometrica}, vol. 70, no. 2, pp. 583-601, 2002.


 \bibitem{technical}
M. Zhang, A. Arafa, E. Wei, R. A. Berry, ``Optimal and quantized mechanism design for timely information acquisition'', [Online]. Available: \url{https://arxiv.org/abs/2006.15751}

\end{thebibliography}
\end{document}